\documentclass[12pt, onecolumn, draftclsnofoot]{IEEEtran}%

\usepackage[T1]{fontenc}
\usepackage{amsmath,amssymb,amsfonts,mathrsfs,bm}
\usepackage{mathtools}
\usepackage{amsthm}
\usepackage{nicefrac}
\usepackage{cite}
\usepackage[shortlabels]{enumitem}
\usepackage{graphicx}
\usepackage{epstopdf}
\usepackage{url}
\usepackage{colortbl}
\usepackage{booktabs}
\usepackage{multirow}
\usepackage[table,dvipsnames]{xcolor}
\usepackage[normalem]{ulem}

\usepackage{array}
\newcolumntype{L}[1]{>{\raggedright\let\newline\\\arraybackslash\hspace{0pt}}m{#1}}
\newcolumntype{C}[1]{>{\centering\let\newline\\\arraybackslash\hspace{0pt}}m{#1}}
\newcolumntype{R}[1]{>{\raggedleft\let\newline\\\arraybackslash\hspace{0pt}}m{#1}}

\makeatletter
\let\MYcaption\@makecaption
\makeatother
\usepackage[font=footnotesize]{subcaption}
\makeatletter
\let\@makecaption\MYcaption
\makeatother

\usepackage{xparse}



\usepackage{glossaries}

\makeatletter
\let\oldgls\gls
\let\oldglspl\glspl

\newcommand\fussy@ifnextchar[3]{%
  \let\reserved@d=#1%
  \def\reserved@a{#2}%
  \def\reserved@b{#3}%
  \futurelet\@let@token\fussy@ifnch}
\def\fussy@ifnch{%
  \ifx\@let@token\reserved@d
    \let\reserved@c\reserved@a 
  \else
    \let\reserved@c\reserved@b
  \fi
 \reserved@c}

\renewcommand{\gls}[1]{%
  \oldgls{#1}\fussy@ifnextchar.{\@checkperiod}{\@}}
\renewcommand{\glspl}[1]{%
  \oldglspl{#1}\fussy@ifnextchar.{\@checkperiod}{\@}}

\newcommand{\@checkperiod}[1]{%
  \ifnum\sfcode`\.=\spacefactor\else#1\fi
}
\makeatother

\newacronym{wrt}{w.r.t.}{with respect to}
\newacronym{RHS}{RHS}{right-hand side}
\newacronym{LHS}{LHS}{left-hand side}
\newacronym{iid}{i.i.d.}{independent and identically distributed}

\usepackage{float}

\ifx\notloadhyperref\undefined
	\ifx\loadbibentry\undefined
		\usepackage[hidelinks,hypertexnames=false]{hyperref} 
	\else
		\usepackage{bibentry}
		\makeatletter\let\saved@bibitem\@bibitem\makeatother
		\usepackage[hidelinks,hypertexnames=false]{hyperref}
		\makeatletter\let\@bibitem\saved@bibitem\makeatother
	\fi
\else
	\ifx\loadbibentry\undefined
		\relax
	\else
		\usepackage{bibentry}
	\fi
\fi

\usepackage[capitalize]{cleveref}
\crefname{equation}{}{}
\Crefname{equation}{}{}
\crefname{claim}{claim}{claims}
\crefname{step}{step}{steps}
\crefname{line}{line}{lines}
\crefname{condition}{condition}{conditions}
\crefname{dmath}{}{}
\crefname{dseries}{}{}
\crefname{dgroup}{}{}

\crefname{Problem}{Problem}{Problems}
\crefformat{Problem}{Problem~(#2#1#3)}
\crefrangeformat{Problem}{Problems~(#3#1#4) to~(#5#2#6)}

\crefname{Theorem}{Theorem}{Theorems}
\crefname{Corollary}{Corollary}{Corollaries}
\crefname{Proposition}{Proposition}{Propositions}
\crefname{Lemma}{Lemma}{Lemmas}
\crefname{Definition}{Definition}{Definitions}
\crefname{Example}{Example}{Examples}
\crefname{Assumption}{Assumption}{Assumptions}
\crefname{Remark}{Remark}{Remarks}
\crefname{Rem}{Remark}{Remarks}
\crefname{remarks}{Remarks}{Remarks}
\crefname{Appendix}{Appendix}{Appendices}
\crefname{Exercise}{Exercise}{Exercises}
\crefname{Theorem_A}{Theorem}{Theorems}
\crefname{Corollary_A}{Corollary}{Corollaries}
\crefname{Proposition_A}{Proposition}{Propositions}
\crefname{Lemma_A}{Lemma}{Lemmas}
\crefname{Definition_A}{Definition}{Definitions}

\usepackage{crossreftools}
\ifx\notloadhyperref\undefined
\else
	\relax
\fi

\usepackage{algorithm,algorithmic}

\ifx\loadbreqn\undefined
	\relax
\else
	\usepackage{breqn} 
\fi

\ifx\renewtheorem\undefined
\ifx\useTheoremCounter\undefined
\newtheorem{Theorem}{Theorem}
\newtheorem{Corollary}{Corollary}
\newtheorem{Proposition}{Proposition}
\newtheorem{Lemma}{Lemma}
\else
\newtheorem{Theorem}{Theorem}

\newtheorem{Proposition}[theorem]{Proposition}
\fi

\newtheorem{Definition}{Definition}
\newtheorem{Example}{Example}

\fi

\theoremstyle{remark}

\theoremstyle{plain}

\newcommand{\scP}{\mathscr{P}}

\DeclareSymbolFont{bsfletters}{OT1}{cmss}{bx}{n}
\DeclareSymbolFont{ssfletters}{OT1}{cmss}{m}{n}
\DeclareMathSymbol{\bsfGamma}{0}{bsfletters}{'000}
\DeclareMathSymbol{\ssfGamma}{0}{ssfletters}{'000}
\DeclareMathSymbol{\bsfDelta}{0}{bsfletters}{'001}
\DeclareMathSymbol{\ssfDelta}{0}{ssfletters}{'001}
\DeclareMathSymbol{\bsfTheta}{0}{bsfletters}{'002}
\DeclareMathSymbol{\ssfTheta}{0}{ssfletters}{'002}
\DeclareMathSymbol{\bsfLambda}{0}{bsfletters}{'003}
\DeclareMathSymbol{\ssfLambda}{0}{ssfletters}{'003}
\DeclareMathSymbol{\bsfXi}{0}{bsfletters}{'004}
\DeclareMathSymbol{\ssfXi}{0}{ssfletters}{'004}
\DeclareMathSymbol{\bsfPi}{0}{bsfletters}{'005}
\DeclareMathSymbol{\ssfPi}{0}{ssfletters}{'005}
\DeclareMathSymbol{\bsfSigma}{0}{bsfletters}{'006}
\DeclareMathSymbol{\ssfSigma}{0}{ssfletters}{'006}
\DeclareMathSymbol{\bsfUpsilon}{0}{bsfletters}{'007}
\DeclareMathSymbol{\ssfUpsilon}{0}{ssfletters}{'007}
\DeclareMathSymbol{\bsfPhi}{0}{bsfletters}{'010}
\DeclareMathSymbol{\ssfPhi}{0}{ssfletters}{'010}
\DeclareMathSymbol{\bsfPsi}{0}{bsfletters}{'011}
\DeclareMathSymbol{\ssfPsi}{0}{ssfletters}{'011}
\DeclareMathSymbol{\bsfOmega}{0}{bsfletters}{'012}
\DeclareMathSymbol{\ssfOmega}{0}{ssfletters}{'012}

\DeclareMathOperator*{\argmin}{arg\,min}

\newcommand{\qednew}{\nobreak \ifvmode \relax \else
      \ifdim\lastskip<1.5em \hskip-\lastskip
      \hskip1.5em plus0em minus0.5em \fi \nobreak
      \vrule height0.75em width0.5em depth0.25em\fi}

\newcommand{\norm}[1]{{\left\lVert{#1}\right\rVert}}

\DeclareDocumentCommand\set{s m t| m}{%
  \IfBooleanTF#1%
	{\left\{\, #2\mathrel{} \IfBooleanTF{#3}{\middle|}{:}\mathrel{}  #4\, \right\}}%
  {\{\, #2 \IfBooleanTF{#3}{\mid}{\mathrel{} : \mathrel{}} #4\, \}}%
}

\DeclareDocumentCommand \ifcond {m m} {%
	{#1} %
	\IfValueT{#2}{\, \middle|\, {#2}}%
}

\DeclareDocumentCommand \P {e{_} g >{\SplitArgument{ 1 }{ @| }}d() g } {%
	\mathbb{P}%
	\IfValueTF{#1}{_{#1}}
		{\IfValueT{#2}{_{#2}}}%
	\IfValueT{#3}{\left(\ifcond#3}%
	\IfValueT{#4}{\, \middle|\, {#4}}%
	\IfValueT{#3}{\right)}%
}

\DeclareDocumentCommand \E {e{_} g >{\SplitArgument{ 1 }{ @| }}o g } {%
	\mathbb{E}%
	\IfValueTF{#1}{_{#1}}
		{\IfValueT{#2}{_{#2}}}%
	\IfValueT{#3}{\left[\ifcond#3}%
	\IfValueT{#4}{\, \middle|\, {#4}}%
	\IfValueT{#3}{\right]}%
}

\definecolor{gray90}{gray}{0.9}

\ifx\nohighlights\undefined

	\newcommand{\msout}[1]{\text{\color{green} \sout{\ensuremath{#1}}}}
	\newcommand{\del}[1]{{\color{green}\ifmmode \msout{#1}\else\sout{#1}\fi}}
\else

	\newcommand{\msout}[1]{#1}
	\newcommand{\del}[1]{#1}
\fi

\newcommand{\hhide}[1]{}

\renewcommand{\figurename}{Fig.}
\newcommand{\figref}[1]{\figurename~\ref{#1}}
\graphicspath{{./Figures/}} 
\newcommand{\includeCroppedPdf}[2][]{%
    \IfFileExists{./Figures/#2-crop.pdf}{}{%
        \immediate\write18{pdfcrop ./Figures/#2 ./Figures/#2-crop.pdf}}%
    \includegraphics[#1]{./Figures/#2-crop.pdf}}

\ifx\diagnoselabel\undefined
	\relax
\else
	\makeatletter
	 \def\@testdef #1#2#3{%
		 \def\reserved@a{#3}\expandafter \ifx \csname #1@#2\endcsname
		\reserved@a  \else
	 \typeout{^^Jlabel #2 changed:^^J%
	 \meaning\reserved@a^^J%
	 \expandafter\meaning\csname #1@#2\endcsname^^J}%
	 \@tempswatrue \fi}
	\makeatother
\fi

\usepackage{tikz-cd}
\usepackage{pgfplots}
\pgfplotsset{compat=1.5}
\begin{document}
\sloppy

\title{Abstract message passing and distributed graph signal processing}
\author{Feng~Ji, Yiqi~Lu, Wee~Peng~Tay, and Edwin~Chong%
\thanks{This work was supported in part by the Singapore Ministry of Education Academic Research Fund Tier 2 grant MOE2018-T2-2-019 and by A*STAR under its RIE2020 Advanced Manufacturing and Engineering (AME) Industry Alignment Fund – Pre Positioning (IAF-PP) (Grant No. A19D6a0053).}%
\thanks{The first three authors are with the School of Electrical and Electronic Engineering, Nanyang Technological University (e-mail: jifeng@ntu.edu.sg, YIQI001@e.ntu.edu.sg, wptay@ntu.edu.sg). E. Chong is with Electrical and Computer Engineering, Colorado State University (e-mail: Edwin.Chong@ColoState.edu).}
}

\maketitle

\begin{abstract}
  Graph signal processing is a framework to handle graph structured data. The fundamental concept is graph shift operator, giving rise to the graph Fourier transform. While the graph Fourier transform is a centralized procedure, distributed graph signal processing algorithms are needed to address challenges such as scalability and privacy. In this paper, we develop a theory of distributed graph signal processing based on the classical notion of message passing. However, we generalize the definition of a message to permit more abstract mathematical objects. The framework provides an alternative point of view that avoids the iterative nature of existing approaches to distributed graph signal processing. Moreover, our framework facilitates investigating theoretical questions such as solubility of distributed problems.
\end{abstract}

\section{Introduction}
Graph signal processing (GSP) has attracted increased attention as it allows us to capture complex correlations in many practical problems. Signals observed in many applications can be modeled as graph signals. Examples include photographs, CMOS sensor images, and readings from sensor networks. GSP has been applied to various problems, including signal recovery, prediction, sampling, and anomaly detection \cite{Shuman2013, San13, San14, Gad14, Moura2015, Don16, Egi17, Isu17, Isu17b, Ort18, JiTay:J19}. 

Given a graph $G$, a \emph{graph signal} on $G$ assigns a value to each node of $G$, resulting in a vector of dimension equal to the size (number of nodes) of the graph. A graph signal is indexed not by time but by the nodes of the graph, called the \emph{graph domain}, capturing irregular domains. A fundamental philosophy of GSP is to perform an orthogonal transformation of the graph domain, called a \emph{graph Fourier transform}. The new domain is usually called the \emph{frequency domain}, analogous to classical Fourier theory. The components of a signal represented in the frequency domain are its \emph{Fourier coefficients}. As in classical signal processing, a graph signal is analyzed by inspecting its Fourier coefficients. An orthogonal transformation is usually obtained via a \emph{graph shift operator} $A$, such as the adjacency matrix or Laplacian of $G$. Under favorable conditions, such as $G$ being undirected, $A$ has an orthonormal eigenbasis and gives rise to the desired transformation. The frequencies are ordered according to size of the eigenvalues.

Basic GSP requires full knowledge of the graph, such as to perform graph Fourier transforms. Such a requirement has shortcomings. For example, graphs can be used to model very large networks such as social networks. Gathering data and network information as well as performing signal processing as above can be costly and time consuming. Moreover, the framework is highly centralized, giving rise to issues such as data privacy. 

On the other hand, some GSP methods do not require the global approach. To give an example, the concept of smoothness corresponds to signals having Fourier coefficients concentrated in the low frequencies. However, to leverage smoothness, one can optimize the quadratic form of total signal variation, which can be done in a decentralized manner. Considerations as such lead to the topic of distributed GSP, including the following efforts. Signal recovery is considered in \cite{Che15, Wan16}. The authors of \cite{Dor15} investigate distributed sparse signal representation with graph spectral dictionaries. Signal reconstruction based on node sampling is studied in \cite{Lor17}. In \cite{Jia19}, the authors consider distributed construction of filter banks. \cite{Ren21} proposes and analyzes a communication efficient distributed optimization framework for general non-convex non-smooth signal processing and machine learning problems under an asynchronous protocol.  

A closely related area of research is distributed optimization. This is a well established field dating back to as early as 1980s \cite{Tsi84,Tsi86, Tsi89}. Though we are not able to give an overview of such a vast topic here, recent survey article \cite{Yan19} contains comprehensive discussions of historical developments and recent advances in distributed optimization. The key component is an iterative procedure that repeats the following: each member of the distributed system solves its own problem using information gathered from its pre-determined ``neighbors'' and share its findings with the neighbors. Theoretical results focus on convergence of such a procedure to the optimal solution. Tables 1--4 in \cite{Yan19} list all the key features of a large collection of algorithms proposed in the literature. 

In terms of methods, the above mentioned works on distributed GSP have spirit common to distributed optimization. In this paper, we are going to consider a different approach. We realize that it is possible that an iterative procedure might be avoided if we tweak the information sharing step with neighbors. Instead of sharing numerical values, we propose sharing elements from a coherent system of vector bundles. The primary example is a smooth function on a domain $D$, whose local incarnation at every point of $D$ is a system of polynomials (by Taylor expansion). The advantage is that more information is contained in such a mathematical object and we only have to deal with each player of the system once. In addition, the framework allows us to leverage numerical invariants such as dimension of manifolds to analyze the problem solving procedure. Thus it is convenient for us to answer theoretical questions such as solubility of a given problem without running the algorithm. In this work, we do not claim that our approach is superior to classical approaches, but we want to provide an alternative point of view, potentially increasing the domain of applicability of GSP. 

The rest of the paper is organized as follows. We formulate the distributed GSP problem in \cref{sec:pro}. In \cref{sec:ner}, we formalize message passing in our setup. In particular, we define what an abstract message is. Moreover, to study solubility questions later on, we introduce the notion of ``local solution''. To handle the abstract notion of message, we need tools from the theory of differentiable manifolds. In \cref{sec:jet}, we introduce jet spaces and Whitney topologies, which are particularly relevant to our discussions. In \cref{sec:sub}, we discuss Morse functions and real analytic functions as the candidates of messages. Both enjoy nice properties that make the procedure of message passing well-behaved. In \cref{sec:sol}, we make use of the theory developed so far to analyze the solubility questions of distributed GSP problems. We propose how to implement our method in practise and present simulation results in \cref{sec:app}, and conclude in \cref{sec:con}.

\section{Problem formulation} \label{sec:pro}

In this section, we describe the setup of the main problem to be studied in the paper, including the notation used throughout the paper.
Let $G=(V,E)$ be an undirected graph, where $V$ is the vertex set and $E$ is the set of edges. Let $t$ be a positive integer, and let $G_i = (V_i,E_i)$, $i\in\{1,\ldots,t\}$ be $t$ subgraphs of $G$ such that $V = \cup_{i=1}^{t} V_i$. For each $i\in\{1,\ldots,t\}$, let $S_i\subset V_i\backslash (\cup_{j\neq i}V_j)$ be a set of nodes contained exclusively in $V_i$ (cf.\ \cref{defn:ato} below). Hence, the $S_i$, $i\in\{1,\ldots,t\}$ are disjoint.

Each node in $S_i$ is understood to be \emph{observable} in the sense that for any signal $x$ on $G_i$, its restriction to $S_i$, denoted $x|_{S_i}$, are known in $G_i$ for the purpose of certain tasks (defined and illustrated by examples below). Such a restriction $x|_{S_i}$ is called an \emph{observation}. Let $c_i=|S_i|$ denote the size of $S_i$ and $S=\cup_{i=1}^{t} S_i$ denote their union, with size $c=|S|=\sum_{i=1}^{t}c_i$ (by disjointness of the $S_i$).

For each $i\in\{1,\ldots,t\}$, recall that graph signals on $G_i$ are functions on the discrete $V_i$, which can be identified with $\mathbb{R}^{|V_i|}$. Let $f_i: \mathbb{R}^{|V_i|} \to \mathbb{R}$ be a convex function on the space of graph signals on $G_i$. Each $f_i$ can be extended to the space of signals on $G$ by composing $f_i$ with the projection $\mathbb{R}^{|V|} \to \mathbb{R}^{|V_i|}$. Let $f: \mathbb{R}^{|V|} \to \mathbb{R}$ denote the sum of these extensions: $f = \sum_{i=1}^{t}f_i$. We are interested in minimizing $f$ and solving certain problems associated with the minimizer. There are  global as well as local versions of these problems, formally introduced next. 

Recall that a manifold $\mathcal{M}$ of dimension $d$ is a Hausdorff topological space that is locally Euclidean of dimension $d$, i.e., every point has an open neighborhood homeomorphic to an open subset of $\mathbb{R}^d$. In Appendix~\ref{sec:fun}, we give a self-contained introduction to the fundamentals of differentiable manifolds.

\begin{Definition} \label{defn:ato}
A \emph{task on $G$} is a function $\tau: \mathbb{R}^{|V|} \to \mathcal{M}$ where $\mathcal{M}$ is a manifold. 

For $U \subset \mathbb{R}^{|S|}$, any graph signal $s \in U$ and minimizer $\hat{x}_s = \argmin_{x|_{S}=s} f(x)$, if $\tau(\hat{x}_s) \in \mathcal{M}$ is well-defined (see explanation below), then we write $\Phi_{\tau}(s)=\tau(\hat{x}_s)$ for the composition of $\tau$ and $\hat{x}_s$. The function $\Phi_{\tau}: U \to \mathcal{M}$ is called the \emph{global problem}.
\end{Definition}


In the definition of $\Phi_{\tau}$, we require $\tau(\hat{x}_s)$ to be well-defined in the following sense. Recognizing that in general $\hat{x}_s$ is not unique, $\tau(\hat{x}_s)$ is \emph{well-defined} if and only if $\tau$ is constant on the set of minimizers $x$ of $f$ subject to $x|_{S}=s$. In particular, if for any $s\in \mathbb{R}^{|S|}$, $f$ has a unique minimizer $x$ such that $x|_{S}=s$, then the domain of $\Phi_{\tau}$ is the full space $\mathbb{R}^{|S|}$. A global problem $\Phi_{\tau}$ is said to be continuous, smooth, etc., if it bears the stated property. Next, we give two examples to illustrate the definitions above.

\begin{Example} \label{eg:ltt}
\begin{enumerate}[1)]
    \item As a first example, let the task $\tau$ be the same as $f: \mathbb{R}^{|V|} \to \mathbb{R}$. Then the global problem $\Phi_{f}$ is well-defined on all of $\mathbb{R}^{|S|}$; i.e., the domain of $\Phi_{f}$ is $\mathbb{R}^{|S|}$.
    \item \label{it:ite} In this second example, we introduce \emph{distributed sampling}, which will be revisited later. Let $z_1,\ldots, z_k \in \mathbb{R}^{|V|}$ be $k$ linearly independent graph signals on $G$ (so $k\leq|V|$). For each $i\in\{1,\ldots,t\}$ and $j\in\{1,\ldots,k\}$, $z_j$ restricted to $V_i$ is denoted ${z_j}|_{V_i}$. Let the function $f_i$ on $G_i$ be given by
    \begin{align*}
        f_i(x) = \min_{(r_1,\ldots,r_k)\in \mathbb{R}^k}
        \norm{\sum_{j=1}^{k}r_j{z_j}|_{V_i} -x}^2,
    \end{align*}
    which is the square distance between $x\in\mathbb{R}^{|V_i|}$ and its orthogonal projection onto the span of $z_1|_{V_i},\ldots, z_k|_{V_i}$. 
    Next, consider the task $\tau: \mathbb{R}^{|V|} \to \mathbb{R}^k$ defined by 
    \begin{align*}
        \tau(x) = \argmin_{(r_1,\ldots,r_k)\in \mathbb{R}^k} \norm{\sum_{j=1}^{k}r_j{z_j} -x}^2,
    \end{align*}
    which is the unique orthogonal resolution of $x$ onto the span of $z_1,\ldots,z_k$. In this case, the codomain manifold $\mathcal{M}$ of $\tau$ is $\mathbb{R}^k$.
    
    For further simplification, let $t=2$ and assume that the two subgraphs $G_1$ and $G_2$ of $G$ have nonempty intersection, i.e., $V_1\cap V_2\neq\emptyset$.
    Suppose that $x$ is a linear combinations of $z_1,\ldots, z_k$. Then, clearly $f_1(x)=f_2(x)=0$. So $\min f = 0$, where $f=f_1+f_2$, and $f_1$ and $f_2$ have been extended to $\mathbb{R}^{|V|}$. 
    
    We now explore conditions under which $\Phi_{\tau}$ is well defined, i.e., for any $s\in\mathbb{R}^{|S|}$, $\tau$ is constant on the set of minimizers of $f$ that agree with $s$ on $S$. To be clear, to say that a signal $x\in\mathbb{R}^{|V|}$ agrees with $s\in\mathbb{R}^{|S|}$ on $S$ means that $x|_{S}=s$. For $\tau$ to be constant on the set of such minimizers, the argmin in the definition of $\tau(x)$ must be the same regardless of the minimizer $x$.
    
    For any minimizer $x$ of $f$, the two vectors of coefficients $(r_1,\ldots,r_k)$ in the definitions of $f_1(x)$ and $f_2(x)$ above are not necessarily the same. So, there are $2k$ decision variables (coefficients). Moreover, these $2k$ coefficients satisfy the following: 
    \begin{enumerate}[(a)]
    \item The two linear combinations of $z_1,\ldots, z_k$ with the two vectors of $k$ coefficients are two signals $s_1$ and $s_2$ that agree on $S$; i.e., their restrictions to $S$ are the same: $s_1|_{S} = s_2|_{S}$. 
    \item The signals $s_1$ and $s_2$ also agree on $V_1\cap V_2$, i.e., $s_1|_{V_1\cap V_2}=s_2|_{V_2\cap V_2}$.
    \end{enumerate}
    To ensure uniqueness of the argmin in the definition of $\tau(x)$ as explained above, we expect that $2k = |V_1\cap V_2| + c$.
\end{enumerate}
\end{Example}

\section{Message passing} \label{sec:ner}

\subsection{Nerve skeleton and message passing}

We want to make use of the message passing paradigm. It is a way of information transfer across a network, and there is nothing new about the process. In this section, we adopt message passing in our setup. Given $G$ and subgraphs $G_i$, $i\in\{1,\ldots,t\}$, we may construct the nerve skeleton $N$ to package information regarding pairwise intersections between different $G_i$'s as follows. Our goal is to solve a global problem in a distributed way. Therefore, to solve the problem on $G_i$ for some $i$, one gathers information from the rest of the graph at the intersections of $G_i$ with the other subgraphs. This motivates us the consider the following.

\begin{Definition}
The \emph{nerve skeleton}\footnote{There is a more general construction called nerve construction giving a simplicial complex. Here we just use its $1$-skeleton, and hence call the resulting graph the nerve skeleton.} $N_G=(V_{N_G}, E_{N_G})$ is an undirected graph of size $t$. Each vertex $g_i$ of $V_{N_G} = \{g_1,\ldots, g_t\}$ corresponds to the subgraph $G_i$. A pair $(g_i,g_j)$ with $i\neq j$ is an edge of $E_{N_G}$ if and only if $G_i\cap G_j \neq \emptyset$.
\end{Definition}

To perform message passing later on, it can be convenient to work with trees (e.g., \cite{Shah11}). Therefore, we are interested in spanning trees of $N$.

\begin{Definition} \label{defn:stn}
Suppose $T\subset N_G$ is a spanning tree. Its complement $T^c$ is the closure of $N_G\backslash T$. If $g \in V_{N_G}$ is considered as a root, then $T_g$ is the unique directed tree on $T$ such that each edge is directed towards $g$, i.e., the head of an edge $(g_i,g_j)$ is $g_j$ if $g_j$ is below $g_i$, where we say that $g_j$ is below $g_i$ if $g_i$ is on the path connecting $g$ and $g_j$.
\end{Definition}

Fixing a root $g$, we use the directed tree $T_g$ to perform message passing, while $T^c$ is used to keep track of connections between intersecting subgraphs not captured by $T$. We now start to discuss message passing.

On $T_g$, consider a directed edge $(g_i,g_j)$ on $T_g$ such that $g_i$ is the tail and $g_j$ is the head, i.e., the direction goes from $g_i$ to $g_j$. Suppose at $g_i$, there is a convex multi-variable function $h_i$. We would like categorize the input variables of the function $h$ as follows:
\begin{enumerate}[1)]
    \item $s_i$: this set of multi-variables corresponds to signals on $S_i$. 
    \item $x_i$: this set of multi-variables corresponds to signals on the union of the set $X_i$ of nodes $V_i\cap V_j$ and $\cup_{(g_i,g_k)\in T^c}(V_i\cap V_k)$. 
    \item $y_i$: this set of multi-variables corresponds to signals on the nodes $Y_i$ belonging only to $g_j$ below $g_i$. In particular, $Y_i$ includes nodes contained exclusively in $V_i$.
    \item $z_i$: this set includes all the remaining variables, use $Z_i$ to denote those coordinates. 
\end{enumerate}

The \emph{message associated with $h_i$ along the edge $(g_i,g_j)$} is 
\begin{align}
    \tilde{h}_i(x_i,z_i) = \min_{y_i} h_i(s_i,x_i,y_i,z_i): \mathbb{R}^{|X_i|+|Z_i|} \to \mathbb{R}.
\end{align}

To explain the domain, we notice that $s_i$ are in fact fixed because we can take observations on $S_i$. Hence, the message $\tilde{h}_i$ is a function on the variables $x_i$ and $z_i$. According to such a definition, a message is not merely a number or a vector, but instead it is a function.

Before describing message passing, we state the following {\bf conventions}, given a function $f: \mathbb{R}^a \to \mathbb{R}$, for convenience: $f$ gives rise to a function, also denoted by $f$, as $f: \mathbb{R}^{a+b} \to \mathbb{R}$, by composing $f$ with the projection $\mathbb{R}^{a+b} \to \mathbb{R}^a$. 

We are ready to describe the \emph{message passing} on $T_g$ with root $g$ as the following procedure:

\begin{enumerate}[S1]
    \item \label{it:sft} Starting from the leaves of $T_g$, each leaf node $g_i$ pass the messages $\tilde{f}_i$ associated with $f_i$ to its only immediate neighbor.
    \item For each node $g_j$ other than $g$, once it receives messages from all edges with $g_j$ as the head node, sum them up and $f_j$ to obtain $h_j$.
    \item The message $\tilde{h}_j$ associated with $h_j$ is passed to $g_k$, along the unique edge $(g_j,g_k)$ such that $g_k$ is the head. 
    \item \label{it:tpt} The procedure terminates at $g$. Take the sum over all the messages received at $g$ including the original convex function at $g$. The resulting function is denoted by $h_g$. 
\end{enumerate}

For later use, we formally extract the key ingredients of the above procedure as follows.

\begin{Definition} \label{defn:wct}
We call the collection of functions $\tilde{f}_i, \tilde{h}_j$ and $h_g$ in \ref{it:sft}$-$\ref{it:tpt} as \emph{messages of the message passing on $T_g$}. The function $h_g$ is the \emph{aggregated message along $T_g$}.
\end{Definition}

We notice that according to the definition, each node is associate with exactly one function as its message.

\begin{figure}
    \centering
	\includegraphics[width=0.3\columnwidth]{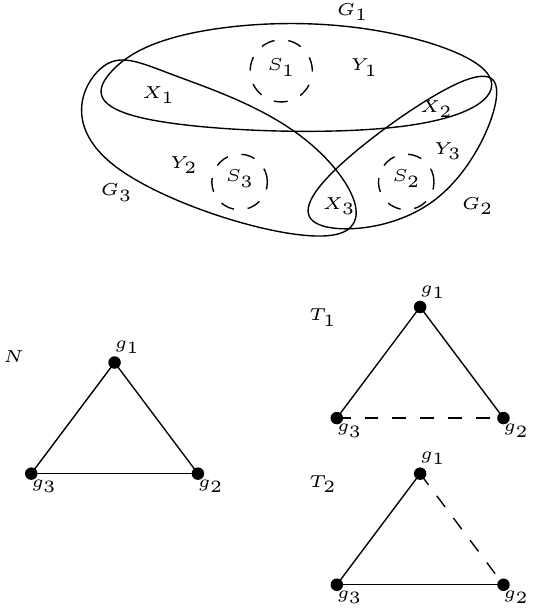}
	\caption{Setup for message passing of \cref{eg:cts}.} \label{fig:dsp1}
\end{figure}

\begin{Example} \label{eg:cts}
For illustration, consider the situation given in \cref{fig:dsp1}. At the top of the figure, we use the Venn diagram to describe the intersection properties of the $3$ subgraphs $G_1, G_2$ and $G_3$. The capital letters label the nodes in respective regions. The nerve skeleton is the complete graph on $3$ nodes as shown on the bottom left. We consider two different spanning trees $T_1$ and $T_2$ shown on the bottom right. Either $T_1^c$ or $T_2^c$ is the single dashed edge. 

For perform message passing on both $T_1$ and $T_2$ with $g_1$ as the root. Let the resulting directed trees be $T_{1,g_1}$ and $T_{2,g_1}$. On $T_{1,g_1}$, messages $\tilde{f}_2(x_2,x_3) = \min_{y_2} f_2(x_2,x_3,y_2,s_2)$ and $\tilde{f}_3(x_1,x_3) = \min_{y_3} f_3(x_1,x_3,y_3,s_3)$ are passed from $g_2, g_3$ to $g_1$ concurrently. At $g_1$, to minimize the aggregated message, we have: $\min_{x_1,x_2,x_3,y_1} f_1 + \tilde{f}_2 + \tilde{f}_3$.

On $T_{2,g_1}$, we first pass the message $\tilde{f}_2(x_2,x_3) = \min_{y_2}f_2(x_2,x_3,y_2,s_2)$ from $g_2$ to $g_3$. At $g_3$, we form the new function $h_3 = f_3 + \tilde{f}_2$, and the message \begin{align*}\tilde{h}_3(x_1,x_2) = \min_{y_3,x_3}(\min_{y_2}f_2(x_2,x_3,y_2,s_2)+ f_3(x_1,x_3,y_3,s_3))\end{align*} is subsequently passed to $g_1$. Finally at $g_1$, minimizing the aggregated message is $\min_{x_1,x_2,y_1} f_1 +\tilde{h}_3$.
\end{Example}

We end this subsection by giving a formal definition of abstract message. Readers unfamiliar with the terminologies may ignore this part, as in the paper we work exclusively with the explicit examples of functions. The terms in the definition will take their concrete forms in \cref{sec:jet}. 

\begin{Definition}
An \emph{abstract message} is an element in the limit of a sequence of vector bundles over a manifold $M$.  
\end{Definition}

We now provide some insights in additional to the concrete examples at hand. Loosely speaking, a vector bundle is a parametrized family of vectors. It contains information about the parameter space, while we can still perform algebraic operations such as taking sums. Therefore, it is a natural choice if we want to generalize numerical or vectorial information. On the other hand, we may also need to work with infinite dimension objects, hence the necessity to consider the limit of a sequence of finite dimensional objects.

\subsection{Local solution}

In \cref{defn:ato}, we have introduced the notion of global problem $\Phi_{\tau}: U \to \mathcal{M}$ given a task $\tau: \mathbb{R}^{|V|} \to \mathcal{M}$, where $U\subset \mathbb{R}^{|S|}$ is the domain of $\Phi_{\tau}$. To understand what we expect from a distributed approach, we fix a spanning tree $T$ of the nerve skeleton $N_G$. 

\begin{Definition} \label{defn:f1w}
For $i\in\{1,\ldots,t\}$, we say that the global problem $\Phi_{\tau}$ can be \emph{solved at $g_i$ via message passing along $T$} if the following holds. Let $h_{g_i}$ be the aggregated message along $T_{g_i}$ with domain $D_i$. There is $\tau_i: D_i \to \mathcal{M}$ such that: for $\hat{x_i} \in \argmin_{x\in D_i} h_{g_i}(x)$, we have $\Phi_{\tau}(s) = \tau_i(\hat{x_i})$.

$\Phi_{\tau}$ can be \emph{solved locally via message passing along $T$} if it can be solved at $g_i$ for every $i\in\{1,\ldots,t\}$.
\end{Definition}

We summarize this definition in the following commutative diagram (with $\hat{x} \in \argmin_x f(x)$):
\[
\begin{tikzcd}[every arrow/.style={draw,mapsto}] 
\{f_1,\ldots,f_t;s\} \arrow{d} \arrow{r} & h_{g_i} \arrow{r} & \hat{x_i} \arrow{d}{\tau_i} \\ \{f,s\} \arrow{r}
& \hat{x} \arrow{r}{\tau} & \Phi_{\tau}(s).
\end{tikzcd}
\]
In the diagram, the top route is via message passing while the bottom route is via solving the global problem. Local solubility requires the existence of the right vertical map $\tau_i$ such that both routes have a ``common destination''.

To give an example, the following result is a generalization of \cref{eg:cts}, which is essentially due to re-arrangement of ordering in taking $\min$. In fact, this result is key in motivating us considering messages as functions. It also gives a concrete example of local solubility to a global problem with minimum requirements on $f_i$, $i\in\{1,\ldots,t\}$. General study of local solubility shall be contained in \cref{sec:sol} below.

\begin{Proposition} \label{prop:ief}
Suppose $\tau = f: \mathbb{R}^{|V|} \to \mathbb{R}$. Then for any $T$ and $i\in\{1,\ldots,t\}$, $\Phi_f$ can be solved locally via message passing along $T$.
\end{Proposition}

\begin{IEEEproof}
We first make some remarks regarding message passing. In general, a sum of convex functions is convex. Minimizing a convex function over a subset of variables is convex on the remaining variable. Therefore, each message is a well defined convex function, as long as so does each $f_i$, $i\in\{1,\ldots,t\}$. 

We prove the result by induction on the size of $T_{g}$. The result is trivially true if $T_{g}$ is a single node. Suppose that $T_g$ contains a leaf node $g_i$, and it is connected by a directed edge $(g_i,g_j)$ to $g_j$. Let $g_{i,1}, \ldots, g_{i,m}$ be the neighbors of $g_i$ in $T^c$, i.e., $\{g_j, g_{i,1}, \ldots, g_{i,m}\}$ are all the neighbors of $g_i$ in $N_G$. During the message passing from $g_i$ to $g_j$, the variables of $f_i$ are re-grouped into $s_i$, $x_i$, $y_i$ and $z_i$, where $x_i, z_i$ accounts for the nodes $V_i \cap (V_j \cup_{i=1}^{m}V_{i,l})$. The message being passed to $V_j$ is thus $\tilde{f_i}(x_i,z_i) = \min_{y_i}f_i(s_i,x_i,y_i,z_i)$. 

For the global problem $\Phi_f$, we want to minimize $f = (f_1 + \ldots + f_{i-1} + f_{i+1} +\ldots + f_t) + f_i$, written as $h+f_i$. We now re-group the variables of $f$ as $x_i,y_i,z_i$ and $r$ where the variables $r$ are disjoint from $x_i,y_i$ and $z_i$. Hence $\Phi_f(s_1,\ldots,s_t) = \min_{x_i,y_i,z_i,r} h(x_i,z_i,r) + f_i(s_i,x_i,y_i,z_i)$. Notice here $s_1,\ldots, s_t$ are fixed numbers at observable nodes. However, $y_i$ is not involved in $h$. Therefore, $\Phi_f(s_1,\ldots, s_t) = \min_{x_i,z_i,r}h(x_i,z_i,r) + \tilde{f}_i(x_i,z_i)$. The right-hand-side is a global problem on the subtree of $T_g$ removing $g_i$, which can be solved locally at $g$ by the induction hypothesis.
\end{IEEEproof}

We end this section by considering the following example where local solution does not exist.

\begin{figure}
    \centering
	\includegraphics[width=0.4\columnwidth]{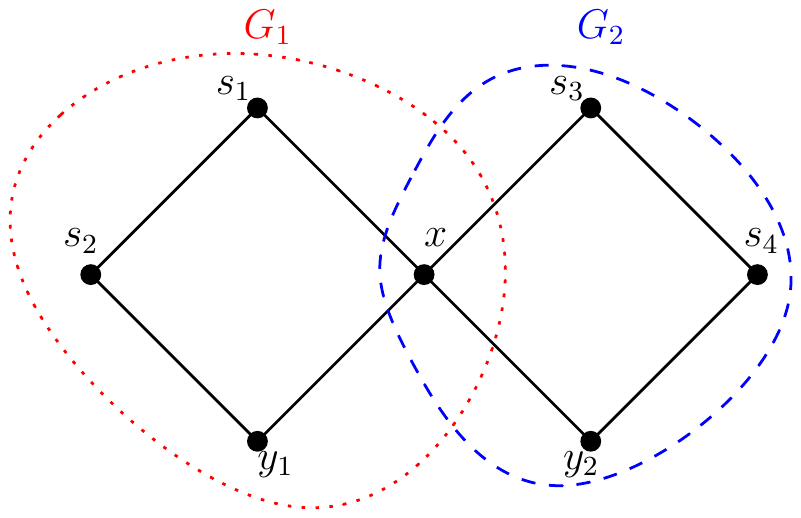}
	\caption{$G_1$ is the square and $G_2$ contains a single edge. They intersect at a single node.} \label{fig:dsp4}
\end{figure}

\begin{Example} \label{eg:tiac}
This is a continuation of \cref{eg:ltt}~\ref{it:ite} and we use the setup, such as choices of $f_i$, $i\in\{1,\ldots,t\}$, stated over there. To be concrete, we assume $t=2$ and the graph $G$ with $2$ subgraphs $G_1=(V_1,E_1)$ and $G_2=(V_2,E_2)$ is shown in \figref{fig:dsp4}. Hence $V_1=\{s_1,s_2,y_1,x\}$, $V_1=\{s_3,s_4,y_2,x\}$ and the intersection $V_1\cap V_2$ contains only $x$. 

To describe $f_1,f_2$ as in \cref{eg:ltt}~\ref{it:ite}, we specify $k=3$. We arrange the vectors $v_1,v_2,v_3$ according to $s_1,s_2,y_1,x,s_3,s_4,y_2$ as:
\begin{align*}
    v_1 = (1, 0, 1, 0, 1, 0, 1)', v_2 =(0, 1, 0, 0, 0, 1, 0)',
     v_3 = (1, 0, 0, 1, 1, 0, 0)'.
\end{align*}
Let the task be $\tau: \mathbb{R}^7 \to \mathbb{R}, (s_1,s_2,y_1,x,s_3,s_4,y_2) \to y_1-y_2$. 

For the global problem $\Phi_{\tau}$, it is in fact smooth as one may verify that $\Phi_{\tau}(s) = s_1-s_3$ for any observation $s = (s_1,s_2,s_3,s_4)$. However, it cannot be solved locally via message passing. To see this, one can show that $\tilde{f}_i$ is the constant $0$ function. One can directly verify this or observe that the components of $v_1,v_2,v_3$ corresponding to $s_1,s_2,x$ are linearly independent. Therefore, the aggregated message at $g_2$ (corresponding to $G_2$) is just $f_2$ itself. Given observations $s_3,s_4$, $\min_{x,y_2} f_2(x,s_3,s_4,y_2)=0$. To minimize $f_2$, the value $y_2$ can be any number. Therefore, we cannot find the $\tau_i$ required in \cref{defn:f1w}.

To give another perspective, as the message is the constant $0$ function, i.e., all the coefficients in the Taylor expansion are $0$. Information on both $s_1,s_2$ are lost, and we have lost ``two degrees of freedom''. Therefore, we are not able to solve the problem at $G_2$. We shall formalize such a point of view in subsequent sections. 
\end{Example}

\section{Jet spaces and Whitney topologies} \label{sec:jet}

In \cref{sec:pro}, we cast the global problem as a function $\Phi_{\tau}: \mathbb{R}^{|S|} \supset U \to M$. We are interested in whether the global problem can be solved locally via message passing in \cref{sec:ner}. During message passing, a message is also an element of a function space. Instead of considering function spaces as sets, which do not have enough structures, our discussions shall revolve around topological structures of smooth function spaces in this section. A self-contained account on background materials is given in Appendix~\ref{sec:fun}.

Recall that a (real) multi-variable function is \emph{smooth} if its partial derivatives of any order exist. For a multi-variable smooth $f$, according to the Taylor's theorem, we can always approximate $f$ around any point by using a polynomial. On the other hand, polynomials of a bounded degree form a finite dimensional vector space. These observations prompts the following \cite{Gol73}.

\begin{Definition}
Let $M,N$ be open subsets of real vector spaces and $C^{\infty}(M,N)$ be the space of smooth functions from $M$ to $N$, i.e., each component is a smooth function on $M$. The \emph{$k$-jet space} $J_p^k(M,N)$ at $p \in M$ is the equivalent classes of $f, g \in C^{\infty}(M,N)$ with: $f \sim g$ if $f$ and $g$ have the same partial derivatives up to $k$-th order at $p$. By convention, equality on the $0$-th order partial derivative means $f(p)=g(p)$. The equivalence class of $f$ is denoted by $f_p$, and the resulting quotient map is $\pi_p^k: C^{\infty}(M,N) \to J_p^k(M,N)$, i.e., $\pi_p^k(f) = f_p$.
\end{Definition}

Apparently, each class of $J_p^k(M,N)$ has a unique representation which is a degree $k$-polynomial. If $M$ is an open subset of $\mathbb{R}^n$ and $N = \mathbb{R}^m$, there is a canonical isomorphism from $J_p^k(M,N)$ to the space of polynomials up to degree $k$: $P^k_{m,n} = (\mathbb{R}[x_1,\ldots,x_n]/(x_1,\ldots,x_n)^{k+1})^m$, the latter is a finite dimensional vector space parametrized by polynomial coefficients. For some simple examples: $\log(1-x) = -x - x^2/2 - x^3/3$ and $\sin(x) = x - x^3/6$ in $J_0^3(\mathbb{R},\mathbb{R})$. 

For each $k\geq 0$, we have the obvious quotient map $J_p^{k+1}(M,N) \to J_p^k(M,N)$, which is also denoted by $\pi_p^{k}$ for convenience (as it is also a quotient map to $J_p^k(M,N)$). 

If we let $p$ vary, then $J^k(M,N)$ is defined as the disjoint union of $J_p^k(M,N)$, namely, $J^k(M,N) = \{(p,f), p\in M, f\in J_p^k(M,N)\}$. As a manifold, $J^k(M,N)$ is identified with the Cartesian product $M \times P^k_{m,n}$. We remark that if $M$ and $N$ are general manifolds, we can still define $J^k(M,N)$ as a manifold by applying the above construction locally using coordinate maps. However, the total space $J^k(M,N)$ is no longer a product space in general. We shall not use the general construction in the sequel and details can be found in \cite{Gol73} Section 2.

\begin{figure}
    \centering
	\includegraphics[width=0.75\columnwidth]{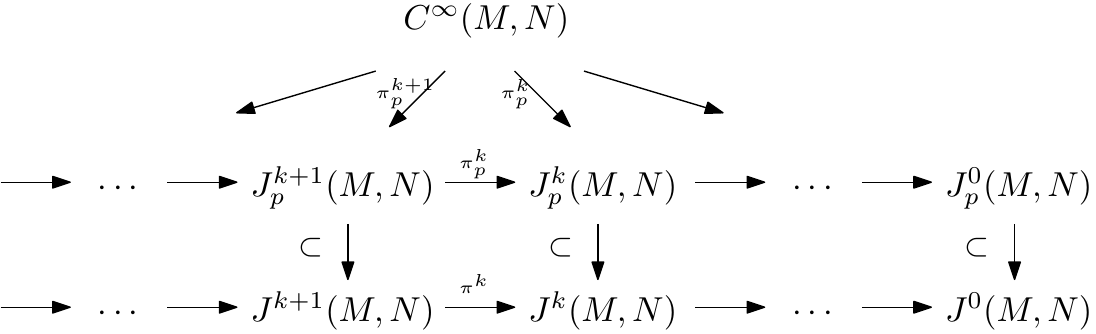}
	\caption{Summary of the relations among the smooth function space and various jet spaces.} \label{fig:dsp2}
\end{figure}

We now use these jet spaces to give $C^{\infty}(M,N)$ topologies, called the \emph{Whitney topologies} \cite{Gol73}. For each $k$ and $f \in C^{\infty}(M,N)$, we have map $\Pi^k(f): M \to J^k(M,N): \Pi^k(f)(p) = (p,f_p)$. Hence, we have the set $\Pi^k(f)(M) \subset J^k(M,N)$ 
\begin{align*}
    \Pi^k(f)(M) = \{(p,f_p) \mid p \in M\}.
\end{align*}
On the other hand, we also have the quotient maps, whose notation inherits from $\pi_p^k$, as $\pi^k: J^{k+1}(M,N) \to J^k(M,N), (p,f) \to (p,f_p)$. For illustration, a summary is given in \figref{fig:dsp2}.

\begin{Definition} \label{def:tbo}
The basis of the \emph{Whitney $C^k$-topology} on $C^{\infty}(M,N)$ is given by: $S^k(U) = \{f \in C^{\infty}(M,N) \mid \Pi^k(f)(M)\subset U\}$, where $U$ is open in $J^k(M,N)$. This means that open sets in $C^{\infty}(M,N)$ are arbitrary unions of sets of the form $S^k(U)$.

Let $W_k$ be the set of open sets given by the $C^k$-topology. The Whitney $C^{\infty}$-topology has basis $W = \cup_{k\geq 0} W_k$.
\end{Definition}

To give an intuition of the Whitney topologies, a sequence of maps $f_n \in C^{\infty}(M,N), n>0$ converges to $f$ in the $C^k$-topology if and only if the following hods: there is a compact subset $K$ of $M$ such that $\Pi^k(f_n): M \to J^k(M,N), n>0$ converges uniformly to $\Pi^k(f)$ on $K$ and all but finitely many $f_n$ agrees with $f$ outside $K$.

Back to the message passing paradigm, we are interested in maps such as $\alpha: U \to C^{\infty}(M,N)$, where $U$ is an open subset of a Euclidean space $\mathbb{R}^d$ and $N = \mathbb{R}$. In turn, composing with $\pi_p^k: C^{\infty}(M,N) \to J_p^k(M,N), p\in M, k\geq 0$ leads to a map $\alpha_p^k: U \to J_p^k(M,N)$, which depends on both $p\in M$ and $k$. To obtain a map independent of $p$, we enlarge the domain and consider 
\begin{align*}
\alpha^k: M\times U \to J^k(M,N), (p,s) \mapsto (p, \alpha_p^k(s)).
\end{align*}
The maps $\alpha^k$ are consistent in the sense that $\alpha^k = \pi^k\circ \alpha^{k+1}$. The construction gives the map $\alpha^k$ between differentiable manifolds $M\times U$ and $J^k(M,N)$. With these maps for different $k$, we want to give $\alpha$ numerical invariants to measure the size of its image.

\begin{Definition} \label{def:sfe}
Suppose for each $k$, there are dense open subsets such $U_k$ of $M \times U$ such that the image $\alpha(U_k)$ of $\alpha^k: U_k \to J^k(M,N)$ is a submanifold of $J^k(M,N)$. Then define $b_{\alpha}^k = \dim (\alpha^k(U_k)) - \dim M$, and $b_{\alpha} = \sup_{k\geq 0} b_{\alpha}^k$. 
\end{Definition}

We use a simple example to illustrate how one may compute something such as $\dim (\alpha^k(U_k))$. Suppose $r_1, r_2$ are two parameters. Then functions $\{(r_1+r_2)^2x^2+(r_1+r_2)x+1 \mid r_1,r_2\in \mathbb{R}\}$ is $1$ dimensional as the coefficient of the degree $1$ term determined uniquely the polynomial function. On the other hand, $\{r_1^2x^2+(r_1+r_2)x+1 \mid r_1,r_2\in \mathbb{R}\}$ is $2$ dimensional as we need to know both the degree $1$ and degree $2$ coefficients to know the polynomial function from the set.

\begin{Proposition} \label{prop:fos}
For open subset $U$ of $\mathbb{R}^d$ and $\alpha: U \to C^{\infty}(M,N)$, we assume that $b_{\alpha}^k,k\geq 0$ are well-defined and let $U_k, k\geq 0$ be given as in \cref{def:sfe}. Then the following holds:
\begin{enumerate}[1)]
    \item If for each $k$, $\alpha^k$ is a submersion at some $p\in U_k$, then $b_{\alpha} \leq d$.
    \item \label{it:tia} Suppose in addition that $\alpha^k$ is a submersion on $U_k$. There is $k_0$ such that for all $k\geq k_0$, $\pi^k: J^{k+1}(M,N) \to J^k(M,N)$ restricts to a diffeomorphism on a dense open subset of $\alpha^{k+1}(U_{k+1})$ to a dense open subset of $\alpha^k(U_k)$. In particular, $ b_{\alpha} = b_{\alpha}^k$ for each $k\geq k_0$.
\end{enumerate}
\end{Proposition}

\begin{IEEEproof}
For 1), as $\alpha^k: U_k \to J^k(M,N)$ is a submersion at $p$, it induces a surjective linear transformation on the tangent spaces of $U_k$ and $\alpha(U_k)$ at $p$ and $\alpha^k(p)$, respectively. We have $\dim (\alpha^k(U_k)) \leq \dim U_k = d+\dim M$. This shows that $b_{\alpha}^k \leq d$ and hence $b_{\alpha}\leq d$.

To prove 2), as the map $\pi^k: J^{k+1}(M,N) \to J^k(M,N)$ is onto, we notice that $b_{\alpha}^k$ is a bounded (by $d$) and non-decreasing sequence,  when $k\to \infty$. By the monotone convergence theorem, $\lim_{k\to \infty} b_{\alpha}^k$ exists and equals to $b_{\alpha}$. As each $b_{\alpha}^k$ is an integer, there is $k_0$ such that $b_{\alpha}^k = b_{\alpha}$ for $k\geq k_0$.

Now for $k\geq k_0$, let $U_{k,k+1} = U_k\cap U_{k+1}$, which is again open dense. Then we have the following relation among various maps and spaces: 
\begin{align*}
    \alpha^k: U_{k,k+1} \stackrel{\alpha^{k+1}}{\to} J^{k+1}(M,N) \stackrel{\pi^k}{\to} J^k(M,N).
\end{align*}
In particular, $\pi^k$ induces a differentiable map from $\alpha^{k+1}(U_{k,k+1})$ to $\alpha^k(U_{k,k+1})$, with both having the same dimension $b_{\alpha}$. On the other hand, the differential $d \pi^k$ is every surjective. Therefore, $d \pi^k$ must be invertible when restricted $\alpha^{k+1}(U_{k,k+1})$. By the inverse function theorem, $\alpha^{k+1}(U_{k,k+1})$ is diffeomorphic to $\alpha^k(U_{k,k+1})$.
\end{IEEEproof}

\section{Messages passing for subfamilies of functions} \label{sec:sub}

The smooth function spaces discussed in \cref{sec:jet} is the playground for us to perform analysis. However, smooth convex functions in general does not behave well under message passing. To analyze local solubility of global problems, we need additional regularities on the functions. On the other hand, we also want the subfamilies contain most familiar functions. In the following, we introduce Morse functions and real analytic functions. 

\subsection{Morse functions}

\begin{Definition}
Suppose $M$ is a differentiable manifold and $f:M \to \mathbb{R}$ is smooth. If at $p\in M$, $df_p=0$, then $p$ is called a \emph{critical point} and $f(p)$ is the \emph{critical value}.

The \emph{Hessian matrix} is $(\partial^2 f/\partial x_i\partial x_j)_{i,j}$. A critical point $p$ is called \emph{non-degenerate} if the Hessian is non-singular at $p$.

The function $f$ is called a \emph{Morse function} if it has no degenerate critical points.  
\end{Definition}

A Morse function $f$ enjoys many nice properties \cite{Mil63}. For example if $M$ is an open subset of $\mathbb{R}^d$, then in an open neighborhood of any critical point $p$, $f$ takes the form $f(p) - x_1^2-\ldots -x_b^2 + x_{b+1}^2 +\ldots + x_d^2$. As a consequence, all the critical points of $f$ are isolated. In particular, if $f$ is also convex, it has a unique minimal point, i.e., $f$ is strictly convex.  

Though it may seen from the definition that Morse functions consist of a restricted subfamily of all smooth functions. However, it is known that they in fact form an open dense subset of all smooth functions, under the $C^2$-topology given in \cref{def:tbo}. This means that Morse functions are omnipresent. 

For later use, we make the following observation regarding Morse functions. 

\begin{Lemma} \label{lem:iia}
Suppose $f(x,y): \mathbb{R}^{m_x+m_y} \to \mathbb{R}$ is a smooth, convex function on sets of multi-variables $x=(x_j)_{1\leq j\leq m_x}$ and $y = (y_k)_{1\leq k\leq m_y}$ of sizes $m_x$ and $m_y$ respectively.
\begin{enumerate}[1)]
\item  If $f_{y_0}(x) = f(x,y_0)$ is a Morse function on $x$, then $\tilde{f}(y) = \min_{x \in \mathbb{R}^{m_x}}f(x,y)$ is a smooth function on an open neighborhood $U_{y_0} \subset \mathbb{R}^{m_y}$ containing $y_0$.

\item \label{it:tso} The set of $y$ such that $f_y=f(\cdot, y)$ being Morse forms an open subset of $\mathbb{R}^{m_y}$.
\end{enumerate}
\end{Lemma}

\begin{IEEEproof}
For 1), to find $\tilde{f}(y)$, we need to solve \begin{align}\frac{\partial f}{\partial x}(x,y) = 0.\label{eq:par}\end{align} If $f_{y_0}(x)$ is a Morse function, then the Hessian of $f_{y_0}$ at $x$ satisfying \cref{eq:par} is non-singular. By the implicit function theorem (c.f.\ \cref{thm:ift}), there is an open subset $U_{y_0}$ containing $y_0$, such that $g(y) = x$ with $(x,y)$ solving Equation~(\ref{eq:par}), is smooth in $y \in U_{y_0}$. Therefore, $\tilde{f}(y) = f(g(y))$ is a smooth function for $y\in U_{y_0}$.

For 2), by 1) and convexity, if $y \in U_{y_0}$, then $f_{y}(x)$ has a unique global minimum. Therefore, it has a unique isolated critical point, which must be non-degenerate. Hence, $f_y$ is a Morse function. 
\end{IEEEproof}

The upshot of this result is that under favorable conditions, such as being Morse on a subset of variables, messages are smooth functions. More concretely, if $g_i$ is a leaf node of a directed tree $T_g$ and $f_i$ satisfies the condition of \cref{lem:iia}, in the initial step of message passing, we obtain a map from $\mathbb{R}^{|S_i|} \to C^{\infty}(M,\mathbb{R})$, where $M$ is a Euclidean space. 

\subsection{Analytic functions}

Another important subfamily of smooth functions are the analytic functions. The subfamily includes many familiar ones such as the polynomials, exponential functions and the trigonometric functions. Formally, a function $f$ is \emph{analytic} in a connected open subset $D$ of a Euclidean space if for each $x\in D$, there is an open neighborhood $D_x\subset D$ of $x$ such that $f$ agrees with its Taylor series expansion about $x$ for any other point in $D_x$. The following observation related to our theme is essentially the identity theorem of analytic functions.

\begin{Lemma} \label{lem:sui}
Suppose $U$ is a connected open subset of $\mathbb{R}^d$ and $\alpha: U \to C^{\infty}(M,N)$ satisfies \cref{prop:fos}~\ref{it:tia} and $k\geq k_0$ as in there. If the image $Im(\alpha)$ contains only analytic functions, then the map $M \times Im(\alpha) \to J^k(M,N), (p,f) \mapsto (p,f_p)$ for $p\in M, f\in Im(\alpha)$ is injective.  
\end{Lemma}

\begin{IEEEproof}
Suppose for analytic functions $f_1,f_2 \in Im(\alpha)$ and $p_1,p_2\in M$, we have $(p_1,f_{1,p})=(p_2,f_{2,p}) \in J^k(M,N)$. Then, $p_1=p_2$ and $f_1$ and $f_2$ have the same partial derivative at $p_1$ up to $k$-th order. As $k\geq k_0$, by \cref{prop:fos}, all the partial derivatives of $f_1$ and $f_2$ at $p_1$ are the same. Therefore, $f_1=f_2$ by the identity theorem \cite{Fri12}.
\end{IEEEproof}

\subsection{Types of message passing}

We come back to message passing in this subsection. Though it is known that Morse functions is dense under $C^2$-topology, the same does not hold for the subspace of smooth convex functions. On the other hand, for optimization, it is favorable to work with convex functions. Therefore, we need a weak notion of ``density'' to deal with the above quagmire. 

Recall that in \cref{sec:jet}, for each $f\in C^{\infty}(M,N)$, we introduce $\Pi^k(f): M \to J^k(M,N)$. For subset $K\subset M$ and open subset $U \subset J^k(M,N)$, we use $S^k(K,U)$ to denote the $f \in C^{\infty}(M,N)$ such that $\Pi^k(f)(K) \subset U$. 
For example, the basis of the $C^k$-topology consists of sets $S^k(U)= S^k(M,U)$. On the other hand, if $K$ is compact, then $S^k(K,U)$ is related to the compact open topology on the functions from $M$ to $J^k(M,N)$. The notion about being ``dense'' is given in a more general form as follows.

\begin{Definition} \label{def:fkf}
Fix $k\geq 0$ and Euclidean spaces $M_i$, $i\in\{1,\ldots,t\}$ as well as $N$. Suppose we are given $W_1\subset W_2 \subset \prod_{i=1}^{t}C^{\infty}(M_i,N)$ and $K = \prod_{i=1}^{t} K_i$ with $K_i\subset M_i$. 

We say that $W_1$ is dense in $W_2$ w.r.t.\ $K$ if the following holds: for every $U = \prod_{i=1}^{t}U_i$ with $U_i$ in $J^2(M_i,N)$ and $(f_i)_{i\in\{1,\ldots,t\}} \in W_2$ with $f_i \in S^k(U_i)=S^k(M_i,U_i)$, there is an $(h_i)_{i\in\{1,\ldots,t\}} \in W_1$ such that $h_i \in S^k(K_i,U_i)$.  
\end{Definition}

For the rest of this section, we work solely with $k=2$. We first consider $t=1$, namely on a single function space $C^{\infty}(M,N)$. Under this definition, dense w.r.t.\ $M$ agrees with the usual notion of dense subset. In general, it is easy to see that if $K_1\subset K_2$, then dense w.r.t.\ $K_2$ implies dense w.r.t.\ $K_1$.

\begin{Proposition} \label{prop:lmm}
Let $M = \mathbb{R}^{m}, m= m_x+m_y$ and $\Delta(M)$ be the space of smooth convex function on multi-variables $x=(x_j)_{1\leq j\leq m_x},y = (y_k)_{1\leq k\leq m_y}$. Define $\Omega(M)$ to be the set of $f \in \Delta(M)$, such that $f_y = f(\cdot,y)$ (on $x$) is Morse for all $y$ in some dense open subset $W_f \subset M_y = \mathbb{R}^{m_y}$. Then for any compact subset $K$ of $M$, $\Omega(M)$ is dense w.r.t.\ $K$ in $\Delta(M)$. 
\end{Proposition}

\begin{IEEEproof}
We first make some general observations regarding the $S^2(K,U)$, with $K$ compact in $M$ and $U$ open in $J^2(M,\mathbb{R})$ containing $\bar{0} = \{(p,0), p\in M\}$. As $J^2(M,\mathbb{R})$ is homeomorphic to $M \times P^2_{m,1}$, for each $p\in M$, we can always find a open set $U_p \in M$ and an open ball $B_p \in P^2_{m,1}$ centered at $0$ with radius $r_p>0$ such that $U_p\times B_p \subset U$. Here, we take note of the fact that the Euclidean space structure of $P^2_{m,1}$ is given by the polynomial coefficients. Therefore, $\cup_{p\in M} U_p\times B_p \subset U$. On the other hand, $K\subset \cup_{p\in K}U_p$. As $K$ is compact, it has a finite subcover, i.e., $K\subset \cup_{i=1}^{l}U_{p_i}$. Let $r = \min \{r_{p_i},i=1,\ldots,l\}$ and $B$ be the open ball in $P^2_{m,1}$ centered at $0$ with radius $r$. Then $K\times B \subset U$. 

The upshot of the discussion is that: any function of the form $h(x,y) = \sum_{1\leq j\leq m_x} a_jx_j^2 + \sum_{1\leq k\leq m_y} b_ky_k^2 \in \Delta(M)$ belongs to $S^2(K,U)$ as long as the non-negative coefficients $a_j,1\leq j\leq m_x$ and $b_k, 1\leq k\leq m_y$ are sufficiently small. This is because for $a_j$'s and $b_k$'s sufficiently small, the coefficients of the Taylor expansion of $h$ belongs to $B$ for each $p \in K$.

We now choose a countable dense subset $\{y_i \in M_y, i\geq 1\}$. By \cref{lem:iia}~\ref{it:tso}, it suffices to show that for any $f \in \Delta(M)$ and any base open neighborhood $S^2(U)$ of the $0$ function, there is an $h\in \Delta(M)\cap S^2(K,U)$ such that $(f+h)_{y_i}$ is Morse for each $i\geq 1$. Here we use the observation that: adding $f$ translates an open neighborhood of $0$ translates to an open neighborhood of $f$, and $f+h$ is viewed as a small perturbation of $f$.

Suppose $h(x,y) = \sum_{1\leq j\leq m_x} a_jx_j^2$ with positive coefficients. Then the tuples $(a_j)_{1\leq j\leq m_x}$ and $i\geq 1$ such that $(f+h)_{y_i}$ is not Morse has Lebesgue measure zero. This is because the sum is not Morse only if some $a_j$ cancels with coefficient of $x_j^2$ in the expansion of $f$ at $y_i$. Such a collection of $(a_j)_{1\leq j\leq m_x}$ has measure zero in $\mathbb{R}^{m_x}$. The set $\{y_i\in \mathbb{R}^{m_y}\}_{i\geq 1}$ is countable. Therefore, there is always $(a_j)_{1\leq j\leq m_x}$, with each component as small as we wish, such that $(f+g)_{y_i}$ is Morse for each $i\geq 1$.
\end{IEEEproof}

We now consider message passing on a directed spanning tree $T_g$ discussed in \cref{sec:ner}. For each $i\in\{1,\ldots,t\}$, we start with a smooth convex functions $f_i \in C^{\infty}(M_i, \mathbb{R})$ where $M_i = \mathbb{R}^{|V_i|}$. The most desirable scenario for us to perform analysis is when all the messages of message passing (\cref{defn:wct}) on $T_g$ are smooth (resp.\ Morse) functions on dense open subsets of the domains, called \emph{smooth (resp.\ Morse) message passing}. Apparently, a Morse message passing is always a smooth message passing. We now discuss how likely they are. 

We notice that the tuple of functions $(f_i)_{i\in\{1,\ldots,t\}}$ belongs to the product space $\Delta = \prod_{i=1}^{t}\Delta(M_i)$. Let $\Omega \subset \Delta$ consist of tuples $(f_i)_{i\in\{1,\ldots,t\}}$ admitting Morse message passing on $T_g$, for any spanning tree $T$ of the nerve skeleton and $g\in G$.

\begin{Theorem} \label{thm:fac}
For any compact subsets $K_i \subset M_i$, $i\in\{1,\ldots,t\}$, we have that $\Omega$ is dense in $\Delta$ w.r.t.\ $K = \prod_{i=1}^{t}K_i$.
\end{Theorem}

\begin{IEEEproof}
The strategy of the proof is similar to that of \cref{prop:lmm}. Namely, we want to modify each $f_i$ by adding a degree $2$ polynomial $q_i$ on the variables of $f_i$, with small positive coefficients. In order to do so, we need to examine the conditions for Morse message passing. As in \cref{prop:lmm}, we want to show that each choice of spanning tree $T$, root node $g$, being Morse message passing on $T_g$ prohibits at most a measure zero set of choices for coefficients. Once this is shown, the rest follows the same argument as in \cref{prop:lmm}.

There are only finitely many choices for $T$ and $g$. We only need to show the above holds for any fixed $T$ and $g$. For any node $g_i$, let $h_i$ denote the function such that its associated message $\tilde{h}_i$ is one of those in \cref{defn:wct}. By definition, $f_i$ is a summand $h_i$. To obtain $\tilde{h}_i$, we need to optimize over the variables associated with $g_j$ below $g_i$ on $T_g$ (c.f.\ \cref{defn:stn}). More precisely, there are two types of such variables: $y$ associated with nodes contained in $G_i$, and $y'$ associated with nodes outside $G_i$. 

To ensure $\tilde{h}_i$ Morse on a dense open subset of its domain, we need to add a positive definite quadratic $q_y$ on $y$ to $f_i$. If a subset of variables $z$ of $y'$ are associated with nodes in $G_j$. We add a quadratic $q_z$ on $z$ to $f_j$, with exactly any one chosen $G_j$ to avoid repetition. It is important to notice that during the message passing until the current stage with $g_i$, we have not performed any optimization over any subset of variables of $y$ and $y'$. Therefore, no quadratic on $z$ is added to $G_j$ until the current stage. As a consequence, to guarantee $\tilde{h}_i$ is Morse, we only need to avoid a measure zero set on the coefficients of quadratic functions $q_y$ and $q_z$'s. 

In subsequent steps of message passing, we do not need to optimize over the variables $y$ and $y'$. Therefore, there are no additional conditions we need to impose on the above mentioned quadratic functions, and this completes the proof.
\end{IEEEproof}

From the proofs, we see that we may modify each of the $f_i$, $i\in\{1,\ldots,t\}$ by adding a positive definite quadratic function on the variables with ``small coefficients'' such that the resulting functions permit smooth (Morse) message passing. Such a procedure could be understood as \emph{regularization}. Moreover, the coefficients can be chosen (uniformly) randomly within a prescribed small domain at $0$.

\section{Solubility results} \label{sec:sol}

In this section, we discuss results on the solubility of a global problem (c.f.\ \cref{sec:pro}) in a distributed way via message passing (c.f.\ \cref{sec:ner}). We give conditions on both the global problem can or cannot be solved locally.

Recall that a global problem takes the form $\Phi_{\tau}: U \to \mathbb{R}^{|V|} \stackrel{\tau}{\to} \mathcal{M}$ (c.f.\ \cref{sec:pro}) for some manifold $\mathcal{M}$. To properly state the results, we assume that $U$ is a connected open subset of $\mathbb{R}^c$ and $\Phi_{\tau}$ is a \emph{smooth surjection}, $c = \sum_{i=1}^{t}c_i$. As in our setup, $c$ is the size of the nodes in $G$ where observation can be made. We remark that requiring $\Phi_{\tau}$ being surjective is not restrictive, for otherwise, we may just consider $\Phi_{\tau}$ as a map from $U$ to its image as long as the image is a manifold. 

\subsection{Individual message passing step} \label{sec:imp}

In this subsection, we examine closely each individual message passing step. For each $k$, there are three spaces involved in the discussion, namely $U$, $J^k(M,N)$ and $C^{\infty}(M,N)$. The first two are manifolds, while $C^{\infty}(M,N)$ in general is not a manifold. Therefore, we want to pass the study of message passing to jet spaces. Suppose $\alpha: U \to C^{\infty}(M,N)$ is given. It induces $\alpha^k: M\times U \to J^k(M,N), (p,s) \mapsto (p,\alpha^k_p(s))$ where $\alpha^k_p(s) = \pi_p^k\Big(\alpha(s) \Big)$ (c.f.\ \cref{sec:jet}). In other words, $\alpha^k$ is the composition $M\times U \stackrel{Id\times \alpha}{\to} M\times C^{\infty}(M,N) \stackrel{\pi^k}{\to} J^k(M,N) $. We make the following assumptions:
\begin{enumerate}[1)]
    \item \cref{prop:fos}~\ref{it:tia} holds and let $k_0$ be as defined there. Moreover, $k\geq k_0$.
    \item The map $\pi^k: \alpha^k(M \times U) \to J^k(M,N), (p, \alpha(s)) \mapsto \Big(p, \pi^k_p(\alpha(s))\Big)$ is injective.
\end{enumerate}
As we have seen in \cref{lem:sui}, the second assumption holds if $\alpha(U)$ contains only analytic functions.

Consider a single instance in message passing $\scP: f \mapsto \tilde{f}$. It extends to a map on pairs $(f,p), p\in M$ by $\scP((f,p)) = (\tilde{f},p')$ with $p'$ is the projection of $p$ to the domain $M'$ of $\tilde{f}$. Suppose we consider a subset $\tilde{M}$ of $M\times C^{\infty}(M,N)$ containing $Im(Id\times \alpha)$, such that $\scP(\tilde{M}) = \tilde{M'}\subset M'\times C^{\infty}(M',N)$. We have the following diagram of maps: 
\[
\begin{tikzcd} 
M\times U \arrow{r}{Id\times \alpha} & \tilde{M} \arrow{r}{\scP} \arrow[swap]{d}{\pi^k} & \tilde{M'} \arrow{d}{\pi^k} \\
& \pi^k(\tilde{M})  \arrow[dashed]{r}{\bar{\scP}} & \pi^k(\tilde{M'}),
\end{tikzcd}
\]
where $\pi^k(\tilde{M})$ and $\pi^k(\tilde{M'})$ belong to $J^k(M,N)$ and $J^k(M',N)$ respectively. Then we can find a set map $\bar{\scP}$ (the dashed arrow) from $\pi^k(\tilde{M})$ to $\pi^k(\tilde{M'})$ making the diagram commute, i.e., $\bar{\scP} \circ \pi^k = \pi^k \circ \scP$. This is because our injectivity assumption on $\pi^k$ guarantees that $\pi^k: \tilde{M} \to \pi^k(\tilde{M})$ is a bijection.

On the other hand, we may also view $Id\times \alpha: M\times U \to M\times C^{\infty}(M,N)$ with a different perspective. Here, we may ignore the first component, being the identity. For each $s\in U$, $\alpha(s): M \to N$. Therefore equivalently, we may interpret this as $F_{\alpha}: M\times U \to N$ by $F_{\alpha}(p,s) = \alpha(s)(p)$. 

\begin{Lemma}
Write the components of $p\in M$ as $p=(x,y)$ such that domain of $M'$ is on $x$. Assume that the following holds:
\begin{enumerate}[1)]
\item \label{it:fmt} $F_{\alpha}: M\times U \to N$ is smooth and Morse on the joint variables $(x,s)$.
\item $\pi^k$ is injective on $\tilde{M}$.
\end{enumerate}
Then $\pi^k \circ \scP \circ \alpha: M\times U \to \pi(\tilde{M'})$ is smooth. Moreover, if $\pi^k\circ \alpha: M\times U \to \pi^k(\tilde{M})$ is a submersion, then $\bar{\scP}$ is smooth.
\end{Lemma}

Notice that $\scP:\tilde{M} \to \tilde{M'}$, where $\tilde{M} \subset M \times C^{\infty}(M,N)$ and $\tilde{M'} \subset M' \times C^{\infty}(M',N)$. Therefore, $\scP$ essentially has two components.

\begin{IEEEproof}
The lemma essentially follows from the implicit function theorem as in \cref{lem:iia}. Given a pair $(p,s) \in M\times U$, its image under $\alpha$ is the pair $(p, f)$, where $f(x,y) = F_{\alpha}(x,y,s)$. Write the variables of $f$ as $p=(x,y)$. The map $\scP$ projects the component $p=(x,y)$ to $p'=x$, which is clearly smooth. We want to show that the partial order derivatives of $\tilde{f}(x) = \min_{y}F_{\alpha}(x,y,s)$ is smooth on $x$ and $s$. By the Morse condition on $F_{\alpha}$, $\tilde{f}(x) = F_{\alpha}(x,g(x,s),s)$ for smooth function $g$. The partial derivatives of $\tilde{f}$ is a polynomial of those of $F_{\alpha}$ evaluated at $(x,g(x,s),s)$ and those of $g$ evaluated at $(x,s)$. Hence, $\tilde{f}$ is smooth on $x$ and $s$ and so are its partial derivatives. 

If $\pi^k\circ \alpha$ is a submersion, it is locally a coordinate projection. Moreover, as $\bar{\scP}\circ \pi^k\circ \alpha = \pi^k\circ \scP \circ \alpha$ is smooth, so is $\bar{\scP}$.
\end{IEEEproof}

\begin{figure}
    \centering
	\includegraphics[width=0.5\columnwidth]{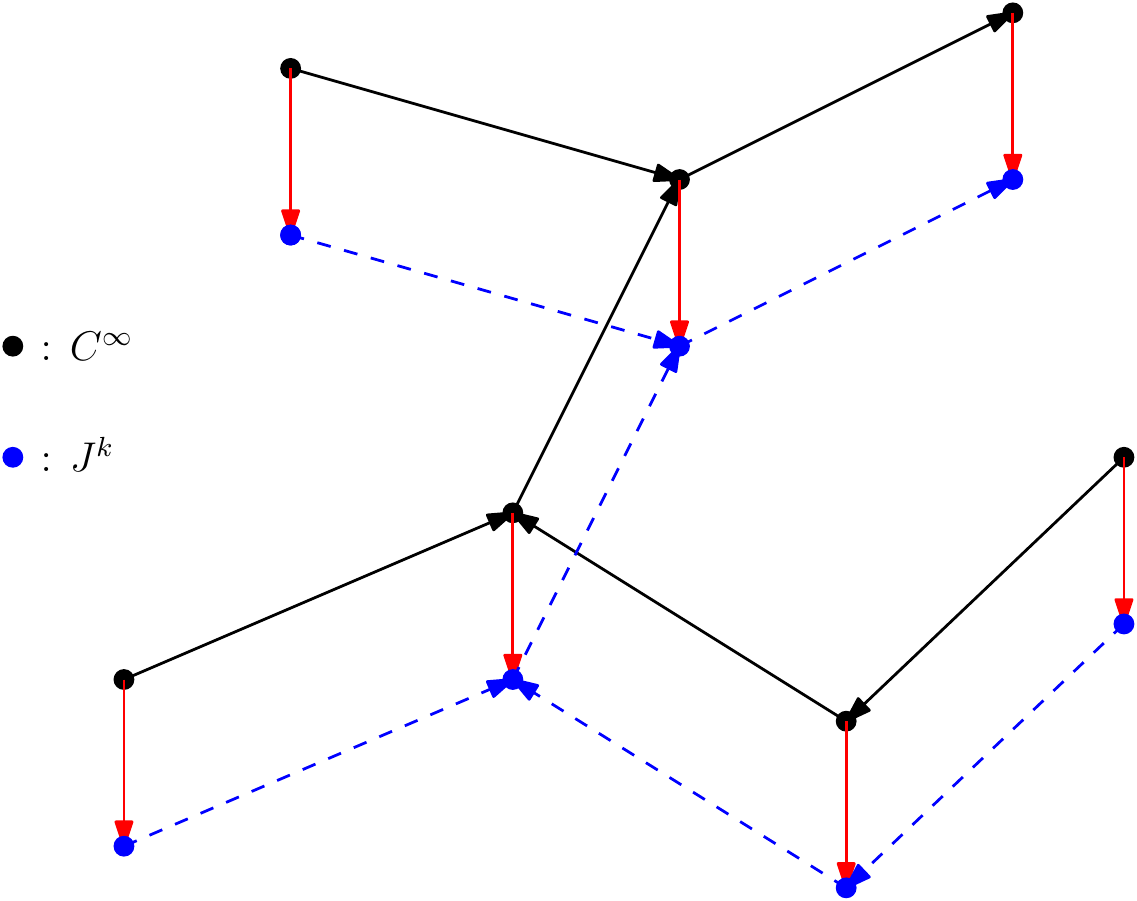}
	\caption{In the top layer, we have the message passing, the spaces are smooth function spaces. In the bottom layer, we have the jet spaces. As we have seen, message passing induces maps in the bottom layer. Study the maps in the bottom layer can be easier as they are between manifolds.} \label{fig:dsp3}
\end{figure}

If we put each individual message passing step together, we obtain a global picture as illustrated in \figref{fig:dsp3}. The main point is that it can usually be difficult or in-explicit to work with smooth functions. However, under favorable conditions, message passing can be viewed as a procedure on the jet spaces, which are Euclidean spaces or more generally manifolds. In doing so, we replace studying functions by studying its derivatives up to certain fixed order. An important advantage is we can now use simple numerical invariants such as dimensions. 

\subsection{Message passing in its entirety}

In this subsection, we are going to state and prove the main result on solubility of a global problem $\Phi_{\tau}$ via message passing. 

\begin{Theorem} \label{thm:ltb}
\begin{enumerate}[1)]
Let $T$ be a spanning tree of $N_G$ and $g_k\in V_{N_G}$.
    \item \label{it:sfa} Suppose $f_1,\ldots, f_t$ admits a Morse message passing. Let $h_k$ be the aggregated message along $T_{g_k}$ (c.f.\ \cref{defn:wct}). Denote the domain of $f=\sum_{i=1}^{t}f_i$ and $h_k$ be $D_f$ and $D_{h_k}$ respectively. Then there is an open dense subset $U$ of $\mathbb{R}^{|S|}, S = \cup_{i=1}^{t}S_i$ such that for any $s=(s_i)_{i=1}^{t}\in U$
    \begin{align*}
    \hat{x} = \argmin_{x \in D_f, x|_S=s} f(x), \text{ and } \hat{y_k} = \argmin_{y \in D_{h_k}, y|_{S_k} = s_k} h_k(y) 
    \end{align*}
    depend smoothly on $s$ and $\hat{y_k} = \hat{x}|_{V_k}$. 
    \item \label{it:fga} For $g_i$ a leaf of $T$ connected to $g_j$, let $M$ be the domain of the message $\hat{f_i}$ from $g_i$ to $g_j$ and $U$ be a dense open subset of $\mathbb{R}^{|S|}$ such that $\hat{f_i}$ is smooth for any $s\in U$. Denote by $\alpha: U \to C^{\infty}(M,\mathbb{R})$. Consider $k_0$ and $b_{\alpha}$ as defined in \cref{def:sfe}, we assume $k\geq k_0$ and $\alpha^k$ is a submersion. If $|S_i|-b_{\alpha} > |S|-\dim \mathcal{M}$, then there does not exist local solution via message passing along $T$.
    
    Moreover, if the condition is verified for one spanning tree $T$, then the (insolubility) conclusion holds for any spanning tree of $N_G$.
\end{enumerate}
\end{Theorem}

\begin{IEEEproof}
\begin{enumerate}[1)]
    \item As we assume that $f_1,\ldots,f_t$ admits a Morse message passing, both $f$ and $h_k$ are smooth Morse functions. Therefore, $\hat{x}$ and $\hat{y_k}$ are uniquely determined. We have shown in \cref{prop:ief} that both $f$ and $h_k$ have the same global minimum. Hence, $\hat{y_k} = \hat{x}|_{V_k}$. It suffices to see that $\hat{x}$ depends smoothly $s$, which follows by applying the implicit function theorem to $f$.
    
    \item If $M$ is the domain of the message from $g_i$ to $g_j$, then we modify the global problem by adding in the identity map on a factor of $M$ as $Id_M\times \Phi_{\tau}: M\times U \to M\times \mathcal{M}$, which remains to be continuous and surjective. Here, $Id_M$ is the identity map on $M$. Suppose under the given conditions that the global problem $\Phi_{\tau}$ has a local solution via message passing along $T$. Then the modified global task $Id_M\times \Phi_{\tau}$ can be decomposed as $M\times U \stackrel{Id_M\times \alpha}{\to} M\times C^{\infty}(M,\mathbb{R})\times U'\to M\times \mathcal{M}$, where $U'$ are on $S\backslash S_i$. The existence of the map $M\times C^{\infty}(M,\mathbb{R})\times U'\to M\times \mathcal{M}$ is due to the assumption on the existence of local solution. Let $\tilde{M}$ be the image of the map $\alpha$. 
    
Consider the diagram of maps 
\[
\begin{tikzcd} 
M\times U \arrow{r}{Id_M\times \alpha} & \tilde{M} \arrow{r} \arrow[swap]{d}{\pi^k\times Id_{U'}}  & M\times \mathcal{M}  \\
& \pi^k(\tilde{M})\times U'  \arrow[dashed]{ur}{\Phi'}.
\end{tikzcd}
\]
As $\pi^k$ is bijective, there is $\Phi':\pi^k(\tilde{M})\times U' \to M\times \mathcal{M}$ making the triangle in the above diagram commute. By our assumptions, $(\pi^k\times Id_{U'})\circ (Id_M\times \alpha)$ is a submersion and $\pi^k$ is a bijection. In particular, $(\pi^k\times Id_{U'})\circ (Id_M\times \alpha)$ is locally a coordinate projection, and $\Phi'$ is differentiable as the top row is the smooth surjection $\Phi_{\tau}$. By Sard's theorem (\cref{thm:sard}), we can find $(p,s) \in M\times U$ such that $Id_M\times \Phi_{\tau}$ is a submersion at $(p,s)$. In view of $b_{\alpha} = \dim(\pi^k(\tilde{M}))-\dim M$ and $\dim U' = |S|-|S_i|$, if $|S_i|-b_{\alpha} > |S|-\dim \mathcal{M}$, we have $\dim(\pi^k(\tilde{M})\times U') < \dim(M\times \mathcal{M})$. This gives a contradiction as $d\Phi_{\tau}$ is surjective at $(p,s)$, while $d\Phi'$ cannot be surjective from a smaller space to a larger space.

The same argument applies to any $T$ as the existence of the diagram does not depend on $T$.
\end{enumerate}
\end{IEEEproof}

Let us summarize what we have found in the results. First of all, \cref{thm:ltb}~\ref{it:sfa} states that when we have a Morse message passing, then the optimizer of the aggregated message agrees exactly with the restriction of the optimizer of the original global problem. If this result is viewed together with \cref{thm:fac}, then we can always add a quadratic regularization term to $f_i$, $i\in\{1,\ldots,t\}$ for the purpose. Of course, we have to pay the price that the solution is not exactly the same as the intended one. On the other hand, \cref{thm:ltb}~\ref{it:fga} gives an explicit condition that local solubility is impossible. The condition is relatively easy to verify as we only need to look at the partial order derivatives of the original functions $f_i$, $i\in\{1,\ldots,t\}$.  

\begin{Example}
We revisit \cref{eg:tiac} to demonstrate the idea of \cref{thm:ltb}~\ref{it:fga}. In that example, $\mathcal{M} = \mathbb{R}$ and its dimension is $1$. The set $S=\{s_1,s_2,s_3,s_4\}$ has size $4$. Hence, $|S| - \dim \mathcal{M}=3$. On the other hand, $|S_1| = 0$. To compute $b_{\alpha}$, as we have seen in \cref{eg:tiac}, the message $\tilde{f}_1$ from $g_1$ to $g_2$ the constant $0$ function, and hence the dimension of the image of $\pi^k$ is $0$. Moreover, the domain of $f_1$ is $2$ dimensional. Therefore, $b_\alpha = 0-2=-2$. Now, we compare that $|S_i|-b_{\alpha}=4>3 = |S|-\dim \mathcal{M}$. By \cref{thm:ltb}~\ref{it:fga}, the global problem cannot be solved locally via message passing. 
\end{Example}

\section{Approximated message passing} \label{sec:app}

So far, we have been mainly focused on theoretical aspects of the message passing scheme. One of the key ingredients is the message, which takes the form $h(x) = \min_{y\in D} h(x,y)$ over variable $y$ over a domain $D$. In practice, it is usually not possible to write down an explicit analytic expression of $h$. In this section, we propose to use a neural network to approximate such a message $h$. 

\begin{figure}	
    \centering
	\includegraphics[width=0.5\columnwidth]{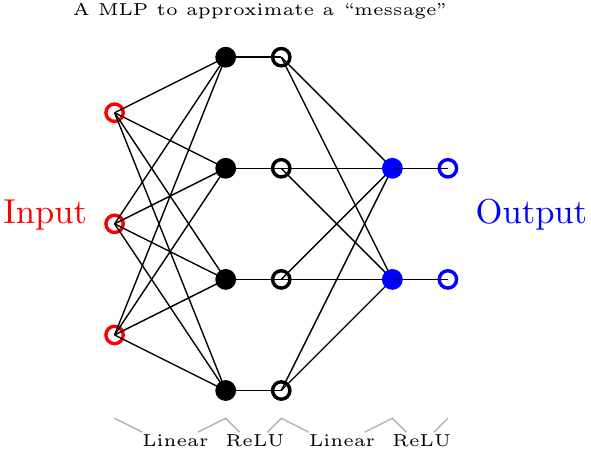}
	\caption{An illustration of MLP.} \label{fig:dsp5}
\end{figure}

Recall that a multilayer perceptron (MLP) (illustrated in \figref{fig:dsp5}) is a function that consists of a concatenation of (hidden) layers, each consist of a composition of a linear transformation and non-linear activation such as the rectifier linear unit (ReLU). For each layer, a finite set of learnable parameters dictates the linear transformation. It is known that a MLP with one hidden layer is enough to approximate any continuous function \cite{Hor89}. Based on this, we propose the \emph{approximated message passing algorithm} as:
\begin{enumerate}[S1']
    \item Construct $N_G$ and fix a spanning tree $T$.
    \item For each $g$, form the directed tree $T_g$. 
    \item Perform message passing following the procedure described in \cref{sec:ner}, with the following modification. 
    \begin{enumerate}[s1]
    \item \label{it:sng} Suppose node $g_j$ is to pass $\tilde{h}_j: D \to \mathbb{R}$ to its neighbor $g_k$. Node $g_j$ will randomly choose samples $x_1,\ldots, x_m \in D$ and compute $y_i = \tilde{h}_j(x_i)$ for each $i\in\{1,\ldots,m\}$. The pairs $\{(x_i,y_i)_{i\in\{1,\ldots,m\}}\}$ are passed to $g_k$. 
    \item Using the received samples $\{(x_i,y_i)_{i\in\{1,\ldots,m\}}\}$, node $g_k$ learns an MLP $\bar{h}_j$ as an approximation of $\tilde{h}_j$. 
    \item One sums up all the approximated messages at $g_k$ and obtain $\tilde{h}_k$. Step~\ref{it:sng} is repeated at $g_k$.
    \end{enumerate}
    \item The aggregated message $h_g$ at $g$ along $T_g$ is optimized locally at $g$.
\end{enumerate}

It is interesting to notice that in the procedure described above, there is only one exchange of information along each edge. This is radically different from many distributed algorithms involving numerous information exchange until convergence. 

For the rest of the section, we present some numerical examples based on the approximated message passing algorithm.

\begin{figure}
    \centering
	\includegraphics[width=0.5\columnwidth]{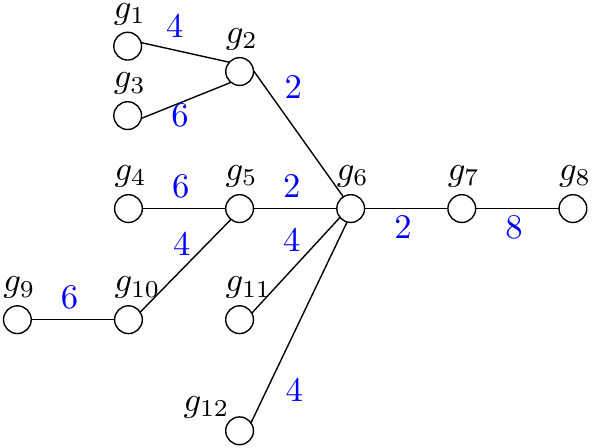}
	\caption{The nerve skeleton.} \label{fig:test1}
\end{figure}

\begin{table}[!htb]
	\caption{A summary of the intersection statistics} \label{tab:1}
	\centering  
	\scalebox{1.2}{
	\begin{tabular}{|l|c|c|c|c|}  
		\hline
		\emph{Subgraph} &  \emph{$|X_i|$} & \emph{$|Y_i|$} & \emph{$|S_i|$} &\emph{$|V_i|$} \\ 
		\hline \hline 
		$G_1$ & 4 & 6 & 12  & 22\\
		\hline
		$G_2$ & 12 & 4 & 10  & 26\\
		\hline
        $G_3$ & 6 & 6 & 14  & 26\\
		\hline
        $G_4$ & 6 & 8 & 12  & 26\\
		\hline
        $G_5$ & 12 & 6 & 10  & 28\\
		\hline
        $G_6$ & 14 & 8 & 6  & 28\\
		\hline
        $G_7$ & 10 & 4 & 12  & 26\\
		\hline
        $G_8$ & 8 & 2 & 14  & 24\\
		\hline
        $G_9$ & 6 & 6 & 12  & 24\\
		\hline
        $G_{10}$ & 10 & 6 & 10  & 26\\
		\hline
        $G_{11}$ & 4 & 8 & 13  & 25\\
		\hline
        $G_{12}$ & 4 & 8 & 14  & 26\\
		\hline
	\end{tabular}}
\end{table}

\begin{table}[!htb]
	\caption{Simulation results ($\%$).} \label{tab:2}
	\centering  
	\scalebox{1.2}{
	\begin{tabular}{|l||c|c|c|c|c|c|}  
		\hline
		\emph{$k$} & 25 & 30 & 35 & 40 & 45 & 50\\ 
		\hline  
		\emph{R($\%$)} &  4.93  &  5.65  & 3.86  &  4.26 &  6.16  & 4.14\\
		\hline
	\end{tabular}}
\end{table}
  
We simulate the distributed sampling in \cref{eg:ltt}~\ref{it:ite}, and follows entirely the setup described in the example. While we are not able to draw the entire graph $G$, we present the nerve skeleton in \figref{fig:test1}. From the figure, we see that there are $12$ subgraphs $G_i=(V_i,E_i)$, $i\in\{1,\ldots,12\}$ of $G$, with $G_i$ corresponds to the node $g_i$. The intersection properties of $G_i$ are given in \cref{tab:1} and \figref{fig:test1}. Recall the notations, for $i\in\{1,\ldots,12\}$, $S_i$ and $Y_i$ are nodes contained exclusively in $V_i$, while $S_i$ are nodes where observations are available. $X_i$ are the nodes in the intersection of $G_i$ with other subgraphs.

For each $i\in\{1,\ldots,12\}$, the function $f_i$ on $G_i$ is
\begin{align*}
    f_i(x) = \min_{r_1,\ldots,r_k} \norm{\sum_{1\leq j\leq k}r_j{z_j}|_{V_i} -x}^2.
\end{align*}
as described in \cref{eg:ltt}\ref{it:ite}. In the simulations, we randomly generated the vectors $z_j, 1\leq j\leq k$. For the parameter $k$, we test the performance of approximated message passing with $k=25,30,\ldots, 50$. For large $k$, each graph has much more unknowns than observable nodes.

We follow the steps of approximated message passing algorithm to get the final approximation of the global optimization. For each simulation, we measure the performance by calculating error between ground truth and estimated optimal value of $f$, and then take the ratio $R$ between error and ground truth. For each $k$, multiple simulations are performed, and the average results are shown in \cref{tab:2}. From the results, we see that although we do not have a perfect result, the approximated message passing yields reasonably good results even when $k$ is large, as our theory expects. 

\begin{figure}
    \centering
	\includegraphics[width=0.5\columnwidth, trim={0 6cm 0 6cm},clip]{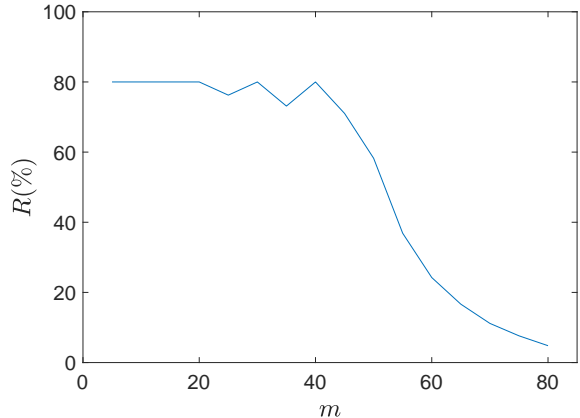}
	\caption{Plot of the error $R$ against the number of samples $m$. In this plot, $R$ being $80\%$ in fact means that the error is larger than $80\%$ and the performance is unstable.} \label{fig:dsp6}
\end{figure}

During message passing, each time we transfer $80$ samples from a node to its neighbor, i.e., $m=80$ in Step~\ref{it:sng}. We further investigate the performance for different $m$ (with fixed choice $k=50$). The results are shown in \figref{fig:dsp6}. We see that the performance starts to show steady improvements from $m=50$ onward, until having the reasonable performance when $m=80$.

\section{Conclusions} \label{sec:con}
In this paper, we propose a framework on distributed graph signal processing. We employ the approach of message passing, and introduce the concept of abstract messages. Our approach provides an alternative point of view of distributed graph signal processing as compare with classical approaches. Moreover, the framework is convenient to analyze theoretical questions such as solubility of distributed problems. Though our work is mainly theoretical, we still present numerical findings to verify the theory. While the topic is distributed GSP, there are still centralized components. Moreover, as we introduce a new notion of message, privacy can be an important concern. In the future, it could be an central topic to investigate.

\appendices

\section{Fundamentals of differentiable manifolds} \label{sec:fun}

In this appendix, we give a self-contained introduction to fundamentals of differentiable manifolds. We shall highlight the results needed for the paper. Readers can further consult textbooks on manifold theory (e.g., \cite{War83}) for thorough discussions of the theory. We start by defining what a topological space is.

\begin{Definition}
A \emph{topology} on a set $X$ is a collection $\mathcal{T}$ of subsets of $X$ s.t.:
\begin{enumerate}[1)]
	\item $\emptyset, X \in \mathcal{T}$.
	\item The union of the elements of any subcollection of $\mathcal{T}$ is in $\mathcal{T}$, i.e., $\mathcal{T}$ is preserved under arbitrary union.
	\item The intersection of the elements of any \emph{finite} subcollection of $\mathcal{T}$ is in $\mathcal{T}$, i.e., $\mathcal{T}$ is preserved under finite intersection.
\end{enumerate}
$X$ with $\mathcal{T}$ is called a \emph{topological space}, and elements of $\mathcal{T}$ are called \emph{open subsets} of $X$.
\end{Definition}

Primary examples of topological spaces include discrete set where each point is open, and Euclidean spaces $\mathbb{R}^n$ where open sets are unions of open balls. In general, a subset $Y$ of $X$ is also a topological space, with an open takes the form $Y\cap U$ where $U$ is open in $X$. 

Not every subset of $X$ is open, but there can be other interesting families. A subset $C$ of $X$ is closed if its complement is open. A subset $D$ is dense in $X$ if it has non-empty intersection with any open subset of $X$. Another key notion is \emph{compact set}: $K$ is compact if a cover of $K$ by open sets (i.e., $K$ is contained in their union) has a finite subcover.

There are many other quantifiers we can associate with topological spaces. We mention two of them here. $X$ is said to be \emph{connected} if it cannot be decomposed into the union of two disjoint non-empty open subsets. It is called \emph{Hausdorff} if for any $x\neq y \in X$, there are disjoint open subset $U_x,U_y$ such that $x\in U_x$ and $y\in U_y$, i.e., one can separate points by using open sets. 

Topological spaces are related by \emph{continuous functions}: $f: X\to Y$ is continuous if the inverse image of any open set is open. A continuous function $f$ is a \emph{homeomorphism} if it has a continuous inverse. 

In addition to being topological spaces related by continuous functions, Euclidean spaces carry more structures, namely differentiable structures. Moreover, we may glue together Euclidean spaces to form more general spaces. We next discuss these ideas and put them together leading to differentiable manifolds.

\begin{Definition} \label{def: fka}
For $k\geq 0$, a function $f: \mathbb{R}^n \to \mathbb{R}$ belongs to $C^k(\mathbb{R}^n,\mathbb{R})$ if $f$ has all continuous partial derivatives up to $k$-th order. The space of smooth functions $C^{\infty}(\mathbb{R}^n,\mathbb{R})$ is $\cap_{k\geq 0} C^{k}(\mathbb{R}^n,\mathbb{R})$, i.e., $f$ is smooth if all partial derivatives exist.

For $f: \mathbb{R}^n \to \mathbb{R}^m$ and $k\geq 0$ or $k=\infty$, it belongs to $C^k(\mathbb{R}^n, \mathbb{R}^m)$ or smooth if each of the $m$-components belongs to $C^k(\mathbb{R}^n,\mathbb{R})$.
\end{Definition}

Taking partial derivative of $f$ at $p$ depends only the value of $f$ at any open set containing $p$. Therefore, the same definition can be used without change if the domain $\mathbb{R}^n$ is replaced by any open subset $U$. If the domain and codomain are clear from context, we may abbreviate $C^k$ or $C^{\infty}$ for convenience. By convention, a function with empty domain belongs to $C^{\infty}$.

A smooth function $f$ is called a \emph{diffeomorphism} if it has a smooth inverse function. We give a more general version of this notion later on once we have introduced differentiable manifolds. Key results regarding differentiable functions are the inverse function theorem and the implicit function theorem \cite{War83, Kau11}.

\begin{Theorem}[The inverse function theorem]
Let $U \subset \mathbb{R}^d$ be open, and let $f: U \to \mathbb{R}^d$ be $C^{\infty}$. If the Jacobian matrix $$\{\frac{\partial r_i\circ f}{\partial r_j}\}_{1\leq i,j\leq d}$$ is non-singular at $r_0\in U$, then there is an open set $V$ with $r_0\in V\subset U$, such that $f(V)$ is open and $f: V\to f(V)$ is a diffeomorphism.
\end{Theorem}

\begin{Theorem}[The implicit function theorem] \label{thm:ift}
Let $U \subset \mathbb{R}^{c-d}\times \mathbb{R}^d$ be open, and let $f: U \to \mathbb{R}^d$ be $C^{\infty}$. We denote the canonical coordinate system on $\mathbb{R}^{c-d}\times \mathbb{R}^d$ by $(r_1,\ldots, r_{c-d}, s_1,\ldots, s_d)$. Suppose that at the point $(r_0,s_0) \in U$, $f(r_0,s_0)=0$; and that the matrix $$Df(r_0,s_0) = \{\frac{\partial f_i}{\partial s_j}|_{(r_0,s_0)}\}_{i,j=1,\ldots, d}$$ is non-singular. Then there exists an open neighborhood $V$ of $r_0$ in $\mathbb{R}^{c-d}$ and an open neighborhood $W$ of $s_0$ in $\mathbb{R}^d$ s.t. $V\times W \subset U$, and there exists a $C^{\infty}$ map $g: V\to W$ s.t.\ for each $(p,q) \in V\times W$: $f(p,q)=0$ iff $q = g(p)$. 
\end{Theorem}

A \emph{locally Euclidean space} $M$ of dimension $d$ is a Hausdorff topological space $M$ for which each point has a neighborhood homeomorphic to an open subset of Euclidean space $\mathbb{R}^d$.

A \emph{differentiable structure} $\mathcal{F}$ on $M$ of class $C^k$ is a collection of coordinate systems $\{(U_{\alpha},\psi_{\alpha})\mid \alpha \in A\}$ with index set $A$ such that the following holds:
\begin{enumerate}[1)]
    \item Each $U_{\alpha}$ is open in $M$ and $M = \cup_{\alpha\in A}U_{\alpha}$.
    \item $\psi_{\alpha}: U_{\alpha} \to V_{\alpha}$ is a homeomorphism, i.e., continuous bijection with a continuous inverse, from $U_{\alpha}$ to an open subset $V_{\alpha}$ of $\mathbb{R}^d$. Each $\psi_{\alpha}$ is called a coordinate map.
    \item If $U_{\alpha}\cap U_{\beta} \neq \emptyset$, then $\psi_{\alpha} \circ \psi_{\beta}^{-1}: W_{\beta} \to V_{\alpha} \in C^k(\mathbb{R}^d,\mathbb{R}^d)$, where $W_{\beta} = \psi_{\beta}(U_{\alpha}\cap U_{\beta})$.
\end{enumerate}
$M$ with such an differentiable structure $\mathcal{F}$ is called \emph{a differentiable manifold of class $C^k$}. We usually consider $k=\infty$ and just call such an $M$ a differentiable manifold. 

The coordinate maps allows us to match open sets of $M$ with those of $\mathbb{R}^d$. As a consequence, we can talk about differentiability of functions on $M$ by pulling back them to a Euclidean space using one of the coordinate maps.

For example, for an open subset $U\subset M$, $f: U\to \mathbb{R}$ is a $C^{\infty}$ function or smooth on $U$ if $f\circ \psi_{\alpha}^{-1}$ is $C^{\infty}$ for each coordinate map $\psi_{\alpha}, \alpha \in A$. 

For differentiable manifolds $M$ and $N$, a continuous map $f: M \to N$ is said to be differentiable of class $C^{\infty}$ or smooth if $\gamma_{\beta} \circ f \circ \psi_{\alpha}^{-1}$ is $C^{\infty}$ for each coordinate $\psi_{\alpha}$ on $M$ and $\gamma_{\beta}$ on $N$. Such a map $f$ is a \emph{diffeomorphism} if $f$ is bijective and $f^{-1}$ is also smooth. 

For each point $p \in M$ contained in $U_{\alpha}$ with coordinate map $\psi_{\alpha}$, a \emph{tangent vector} is a linear map on the space of smooth functions satisfying the Leibniz rule. The tangent space $T_pM$ is the vector space of tangent vectors spanned by the ``partial derivatives'' $\frac{\partial}{\partial x_i }|_p$: 
\begin{align*}
    \frac{\partial}{\partial x_i}|_p(f) = \frac{\partial (f\circ \psi_{\alpha}^{-1})}{\partial r_i}|_{\psi_{\alpha}(p)},\ i\in\{1,\ldots,d\},
\end{align*}
where $f \in C^{\infty}(M,\mathbb{R})$ and $r_i$ is the $i$-th coordinate function of $\mathbb{R}^d$. 

Let $f: M \to N$ be a smooth function and $p\in M$. The \emph{differential of $f$} at $p$ is the linear map $df_p: T_pM \to T_{f(p)}N$ by $$df_p(v)(h) = v(h\circ f),$$ for $v \in T_pM$ and $g \in C^{\infty}(N,\mathbb{R})$. Locally at $p$, the differential $df_p$ gives a linear approximation of $f$. Many properties of $f$ can studied through $df_p$, which is more accessible being a linear transformation. To demonstrate, we define what immersion, imbedding and submersions are. 

The smooth function $f$ is called an \emph{immersion} if $df_p$ is non-singular for every $p\in M$. The manifold $M$ with $f$ is a \emph{submanifold} of $N$ if $f$ is both injective and an immersion. One should take note here that we are not making a repetition of conditions as being being an immersion refers to injectivity of the differential at each point, but not the injectivity of $f$ itself. An even stronger notion is that of \emph{imbedding}: $M$ is a submanifold and $f:M \to f(M)$ is a homeomorphism. These notions are all nonequivalent, and \cite{War83} 1.28 contains concrete examples for this. On the other hand, $f$ is a \emph{submersion} if $df_p$ is surjective for every $p\in M$. A consequence of being submersion is that for every $p\in M$, there is an open neighborhood $U_p$ of $p$ diffeomorphic to a Euclidean open set such that $f$ restricts to a coordinate projection on $U_p$. In particular, $f$ is an open map from $f$ to $f(M)$, i.e., the image of open set is open. Though not every smooth surjective map is a submersion (e.g., $f: \mathbb{R} \to \mathbb{R}, x \to x^3$ is not a submersion at the single point $x=0$), Sard's theorem \cite{Sar42} states that for most points $f$ is a submersion. More precisely, we have the following special case of Sard's theorem:

\begin{Theorem} \label{thm:sard}
Let $f: M \to N$ be a smooth surjection between differentiable manifolds. If $X\subset M$ consists points where $f$ is not a submersion, then $f(X)$ has Lebesgue measure $0$ in $N$.  
\end{Theorem}

\bibliographystyle{IEEEtran}
\bibliography{IEEEabrv,StringDefinitions,reference}

\end{document}